\documentclass[structabstract]{aa}
\usepackage{graphicx}
\usepackage{txfonts}
\input{epsf}
\usepackage{rotating}
\newcommand{\mic}{\,$\mu$m}

\newcommand{\dg}{$^{\circ}$}

\newcommand{\htwo}{H$_2$ 2.122\,$\mu$m}

\begin{document}
  
\title{A census of molecular hydrogen outflows and their sources
             along the Orion~A molecular ridge}
  
\subtitle{Characteristics and overall distribution}
  
\authorrunning{Davis et al.}

\author{C.J.~Davis\inst{1}, 
        D.~Froebrich\inst{2},
        T.~Stanke\inst{3},
        S.T.~Megeath\inst{4},
        M.S.N.~Kumar\inst{5},
        A. Adamson\inst{1},
        J.~Eisl\"offel\inst{6},
        R.~Gredel\inst{7},
        T.~Khanzadyan\inst{8},
        P.~Lucas\inst{9},
        M.D.~Smith\inst{2} \and 
        W.P.~Varricatt\inst{1}
       }

\institute{
Joint Astronomy Centre, 660 North A'oh\={o}k\={u} Place,
    University Park, Hilo, Hawaii 96720, U.S.A. \\
    \email{c.davis@jach.hawaii.edu}
  \and
Centre for Astrophysics \& Planetary Science,
    School of Physical Sciences, University of Kent,
    Canterbury CT2 7NR, U.K.
  \and
European Southern Observatory, Garching, Germany 
  \and
Department of Physics and Astronomy, University of Toledo, Toledo, 
    OH 43606-3390, U.S.A.
  \and
Centro de Astrofisica da Universidade do Porto,
    Rua das Estrelas s/n 4150-762 Porto, Portugal 
  \and
Th\"uringer Landessternwarte Tautenburg, Sternwarte 5, 
    D-07778 Tautenburg, Germany 
  \and
Max Plank Institute f\"ur Astronomie, K\"onigstuhl 17, 
    D-69117 Heidelberg, Germany 
  \and
Centre for Astronomy, Department of Experimental Physics, National
    University of Ireland, Galway, Ireland
  \and
Centre for Astrophysics Research, Science \& Technology Research Institute, 
    University of Hertfordshire, College Lane, Hatfield AL10 9AB, U.K. 
}

\date{Received: October 1, 2008.  Accepted: ...}
  
\abstract{

{{\em Aims.} A census of molecular hydrogen flows across the entire Orion~A
Giant Molecular Cloud is sought.  With this paper we aim to associate each
flow with its progenitor and associated molecular core, so that the
characteristics of the outflows and outflow sources can be
established.}

{{\em Methods}. We present wide-field near-infrared images of Orion~A,
obtained with the Wide Field Camera, WFCAM, on the United Kingdom
Infrared Telescope. Broad-band K and narrow-band H$_2$ 1-0S(1) images
of a contiguous $\sim$8 square degree region are compared to mid-IR
photometry from the Spitzer Space Telescope and (sub)millimetre
dust-continuum maps obtained with the MAMBO and SCUBA bolometer
arrays.  Using previously-published H$_2$ images, we also measure
proper motions for H$_2$ features in 33 outflows, and use these data
to help associate flows with existing sources and/or dust cores.}

{{\em Results}. Together these data give a detailed picture of
dynamical star formation across this extensive region.  We increase
the number of known H$_2$ outflows to 116.  A total of 111 H$_2$
flows were observed with Spitzer; outflow sources are identified for
72 of them (12 more H$_2$ flows have tentative progenitors).  The
MAMBO 1200\mic\ maps cover 97 H$_2$ flows; 57 of them (59\%) are associated
with Spitzer sources and either dust cores or extended 1200\mic\ emission.
The H$_2$ jets are widely distributed and randomly orientated; the
jets do not appear to be orthogonal to large-scale filaments or even
to the small-scale cores associated with the outflow sources (at least
when traced with the 11\arcsec\ resolution of the 1200\mic\ MAMBO
observations).  Moreover, H$_2$ jet lengths ($L$) and opening angles
($\theta$) are not obviously correlated with indicators of outflow
source age -- source spectral index, $\alpha$ (measured from
mid-IR photometry), or (sub)millimetre core flux.  It seems clear that
excitation requirements limit the usefulness of H$_2$ as a tracer of
$L$ and $\theta$ (though jet position angles are well defined).}

{{\em Conclusions}. We demonstrate that H$_2$ jet sources are
predominantly protostellar sources with flat or positive
near-to-mid-IR spectral indices, rather than disk-excess (or T Tauri)
stars.  Most protostars associated with molecular cores drive H$_2$
outflows.  However, not all molecular cores are associated with
protostars or H$_2$ jets.  On statistical grounds, the H$_2$ jet phase
may be marginally shorter than the protostellar phase, though must be
considerably (by an order of magnitude) shorter than the prestellar
phase.  In terms of range and mean value of $\alpha$, H$_2$ jet
sources are indistinguishable from protostars. The spread in $\alpha$
observed for both protostars and H$_2$ outflow sources is probably a
function of inclination angle as much as source age. The few true
protostars without H$_2$ jets are almost certainly more evolved than
their H$_2$-jet-driving counterparts, although these later
stages of protostellar evolution (as the source transitions to being a
``disk excess'' source) must be very brief, since a large fraction of
protostars do drive H$_2$ flows. We also find that protostars
that power molecular outflows are no more (nor no less) clustered than
protostars that do not. This suggests that the H$_2$
emission regions in jets and outflows from young stars weaken and fade very
quickly, before the source evolves from protostar to pre-main-sequence
star, and on time-scales much shorter than those associated with the
T~Tauri phase, the Herbig-Haro jet phase, and the dispersal of young
stellar objects. }


}
  
\maketitle

%%%%%%%%%%%%%%%%%%%%%%%%%%%%%%%%%%%%%%%%%%%%%%%%%%%%%%%%%%%
%%%%%%%%%%    FIGURES - OVERVIEW      %%%%%%%%%%%%%%%%%%%%%
%%%%%%%%%%%%%%%%%%%%%%%%%%%%%%%%%%%%%%%%%%%%%%%%%%%%%%%%%%%

%%%doublecolumn figure
\begin{figure*}
\centering
  \epsfxsize=17.5cm
  \epsfbox{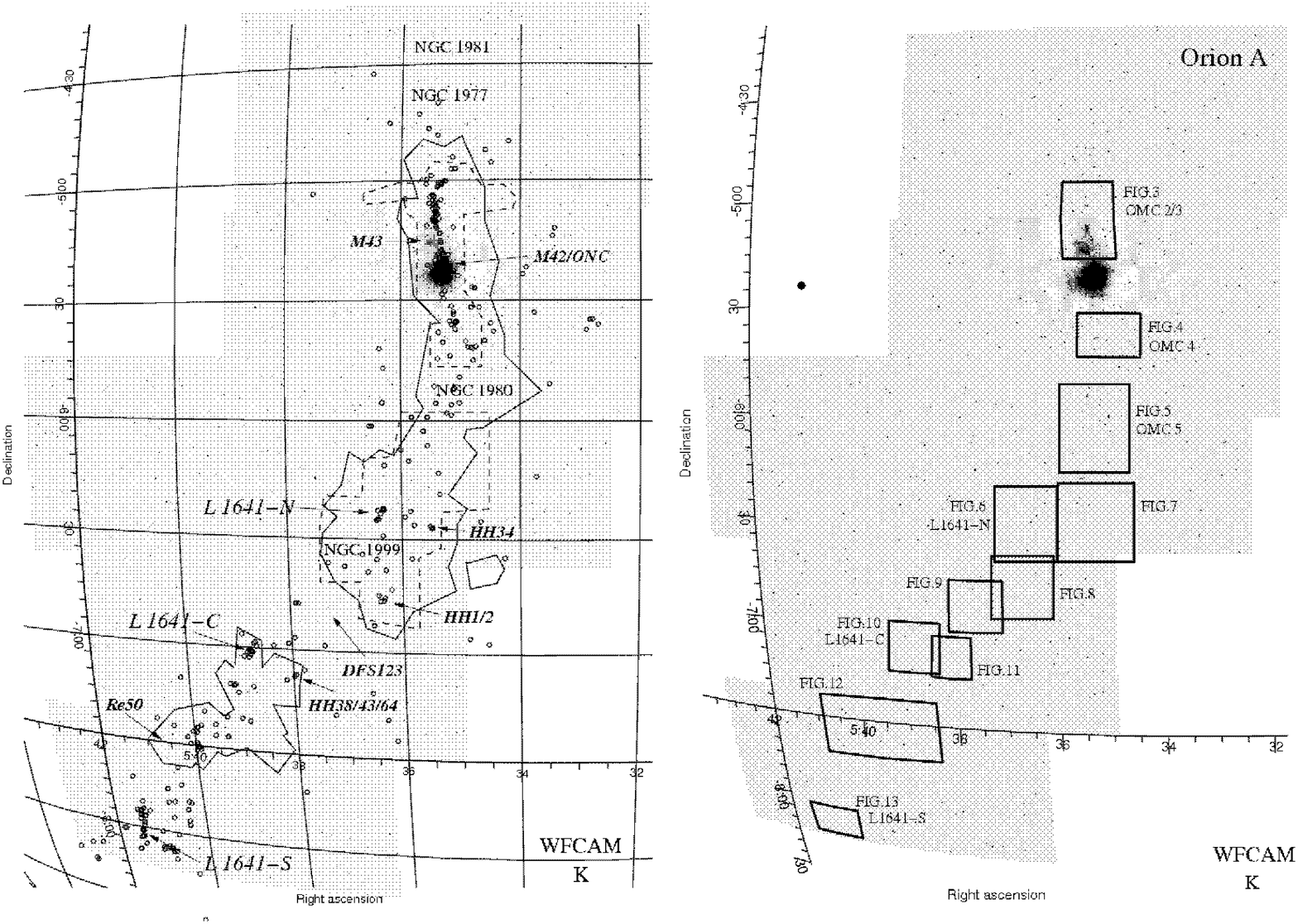}
%\vspace*{0.3cm}
\caption {Our near-IR K-band mosaic of Orion~A.  In the left-hand panel,
the positions of candidate protostars identified from Spitzer
photometry are marked with circles; the areas outlined with full lines
have been mapped at 1200\mic\ (Stanke et al., in prep.); the areas
marked with dashed lines have been mapped at 850\mic\ (\cite{nut07}
2007). In the right-hand panel the regions shown in detail in
Figs.~\ref{scu1} to \ref{scu11} are indicated with boxes. }
\label{over}
\end{figure*}

%%%%%%%%%%%%%%%%%%%%%%%%%%%%%%%%%%%%%%%%%%%%%%%%%%%%%%%%%%%%%%%%%
%%%%%%%%%%%%%%%%%%%%%%%%%%%%%%%%%%%%%%%%%%%%%%%%%%%%%%%%%%%%%%%%%

\section{Introduction}

The southern part of the Orion constellation encompasses the Orion~A
and B Giant Molecular Clouds (GMCs), numerous compact and
intermediate-sized molecular cores, low and high mass young stars
(including a massive star/young OB cluster), stars at varying
evolutionary stages, and dozens of optical Herbig-Haro (HH) objects
and molecular outflows (see \cite{pet08} (2008) and \cite{all08}
(2008) for reviews).  In molecular line maps the complex extends
roughly northwest-southeast, parallel with the Galactic plane, over
about 13$^{\circ}$ (\cite{kut77} 1977; \cite{sak94} 1994).  Overall,
the Orion A and B GMCs represent one of the richest star forming
regions known.

In extensive regions like Orion, jets and outflows can be used as
sign-posts of on-going star formation.  An abundance of jets points to
active accretion and a young stellar population; a paucity of 
molecular outflows, in a region where near- and mid-IR photometry
still indicate a sizable population of sources with reddening and
excess, suggests a more evolved region with a fraction of
pre-main-sequence stars (T Tauri and Herbig Ae/Be stars) observed
``edge-on'' through their circumstellar disks.

By using \htwo\ emission as a tracer of jets and outflows, wide-field
narrow-band images may be used to pin-point the locations of the
youngest sources.  Moreover, they allow one to take a statistical
approach when considering questions about the distribution of Class
0/I protostars and Class II/III Young Stellar Objects (YSOs), the
interaction of these forming stars with their surrounding environment,
the overall star formation efficiency, and the evolution of the region
as a whole (e.g. \cite{sta00} 2000; \cite{sta02} 2002; \cite{dav07}
2007; \cite{kum07} 2007; \cite{dav08} 2008).

With the wide-field near-IR camera WFCAM at the United Kingdom
Infrared Telescope (UKIRT) we have secured homogeneously deep,
sub-arcsecond-resolution \htwo\ and broad-band K images over most of
the Orion~A GMC.  Our $\sim$8 square degree mosaics encompass the
the molecular clouds known as OMC~2 and OMC~3, 
the Orion Nebula Cluster (ONC -- also known as M~42), and an
abundance of well-known low mass star forming cores, HH
objects and bipolar outflows spread throughout the Lynds dark cloud
L~1641 (HH~1/2, HH~33/40, HH~34, HH~38/43, etc.).

Our goals with these observations were to:
(1) extend the maps of Stanke et al. (2002; hereafter Sta02) to give more
complete coverage of Orion~A, to search for new H$_2$ flows, and to
better trace the true extent of known H$_2$ flows by searching for
emission off the main molecular ridge;
(2) provide a second epoch for proper motion studies, and
(3) compare the observations with extensive (sub)millimetre and mid-IR
Spitzer observations, so that H$_2$ jets and outflows could be associated
with molecular cores and/or embedded protostars, the latter being
identified from mid-IR photometry.  We utilise the 850\mic\ SCUBA data
published by \cite{nut07} (2007), the more extensive 1200\mic\ MAMBO
observations of Stanke et al. (in prep.), and the recent Spitzer
observations of Megeath et al. (in prep.).

This paper is structured as follows: in Sect.3 we briefly discuss
wide-field images of regions of interest spread throughout Orion~A and
give statistical information pertaining to the overall population of
outflows. In Tables~\ref{smz} and \ref{dfs} we list the H$_2$ flows
catalogued by Sta02 and the newly identified H$_2$ flows from this
paper, respectively: in both tables we list the likely outflow source
(if known), the dense cores that coincide with the source, and any HH
objects that are associated with the flow. In Sect.4. we discuss the
region as a whole, drawing statistical information from the sample of
flows and associated sources.  In Appendix A we briefly discuss the
full sample of molecular H$_2$ outflows, presenting
continuum-subtracted \htwo\ images of the newly-observed flows. In
Appendix B we discuss our proper motion measurements for outflows
observed here and in the earlier work of Sta02.

%%%%%%%%%%%%%%%%%%%%%%%%%%%%%%%%%%%%%%%%%%%%%%%%%%%%%
%%%%%%%%%%    FIGURES    COLOR  %%%%%%%%%%%%%%%%%%%%%
%%%%%%%%%%%%%%%%%%%%%%%%%%%%%%%%%%%%%%%%%%%%%%%%%%%%%

%%%doublecolumn figure
\begin{figure*}
\centering
%  \epsfxsize=17cm
%  \epsfbox{davis-fg2.ps}
\vspace*{5.0cm}
{\bf Colour high-def picture available from http://www.jach.hawaii.edu/$\sim$cdavis/
\\
Or see JPEG figure 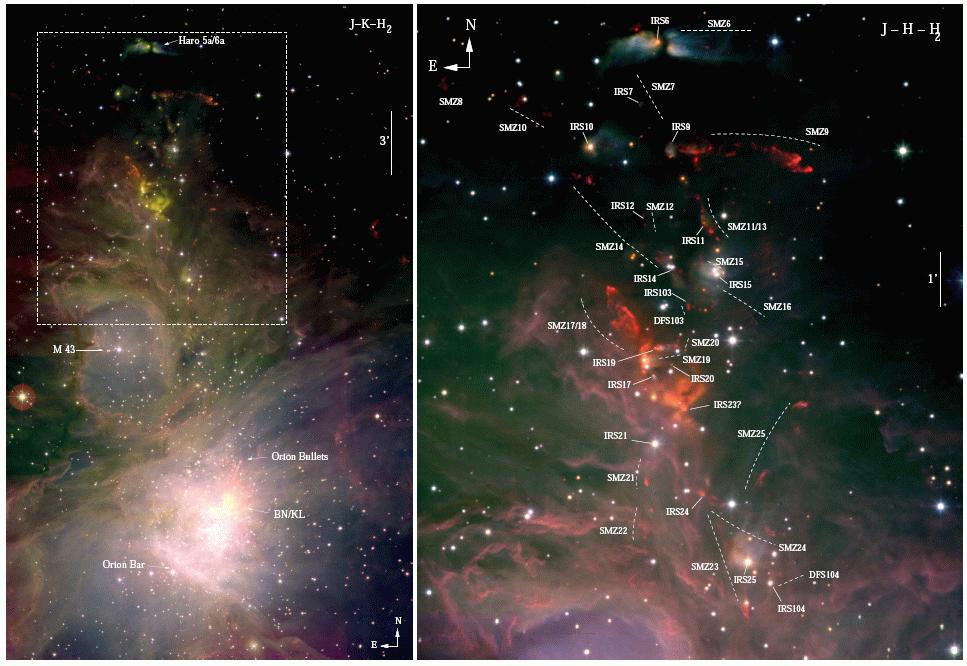 with this astro-ph submission.
}
\vspace*{5.0cm}
\caption {Left: colour image of OMC~1/2/3 composed of broad-band J, K
and narrow-band \htwo\ observations.  The data have been stretched
logarithmically.  Right: colour J, H and narrow-band \htwo\ composite
of the region between Haro 5a/6a and M~43 (OMC~2/3).  The data have
again been stretched logarithmically.  H$_2$ flows are labelled with
``SMZ'' or ``DFS'' numbers; outflow sources are labelled with ``IRS''
numbers. The dashed box in the left-hand image marks the area covered
by the right-hand image.}
\label{color}
\end{figure*}

%%%%%%%%%%%%%%%%%%%%%%%%%%%%%%%%%%%%%%%%%%%%%%%%%%%%%%%%%%%%%%%%%
%%%%%%%%%%%%%%%%%%%%%%%%%%%%%%%%%%%%%%%%%%%%%%%%%%%%%%%%%%%%%%%%%

\section{Observations}

\subsection{Near-IR imaging}

Broad-band J, H, K and narrow-band \htwo\ images of a 1.5\dg
$\times$1.5\dg\ field centred on M~42/M~43 were obtained during
instrument commissioning on 24 November 2004 (filter
characteristics are described in \cite{hew06} 2006).  Broad-band K and
narrow-band H$_2$ images of a $\sim$8 square degree field were later
secured during Director's Discretionary Time (DDT), on 15-16 December
2005. K and H$_2$ images of a single 0.75 square degree segment centred
on the Orion nebula were also obtained during Service observing time
on 12 February 2007. The commissioning data have only been used to
construct the colour images in Fig.~\ref{color}; the
later observations, which were secured under slightly better observing
conditions, and when the instrument was better characterised, have
been used for all other figures and analysis.  The service data have
been used to fill in a gap in the DDT data, where bright stars and
saturation affects produced poor results in the processing.

%%%%%%%%%%%%%%%%%%%%%%%%%%%%%%%%%%%%%%%%%%%%%%%%%%%%%
%%%%%%%%%%    FIGURES H2 B&W with MAMBO  %%%%%%%%%%%%
%%%%%%%%%%%%%%%%%%%%%%%%%%%%%%%%%%%%%%%%%%%%%%%%%%%%%

%%%doublecolumn figure
\begin{figure*}
\centering
  \epsfxsize=17.0cm
  \epsfbox{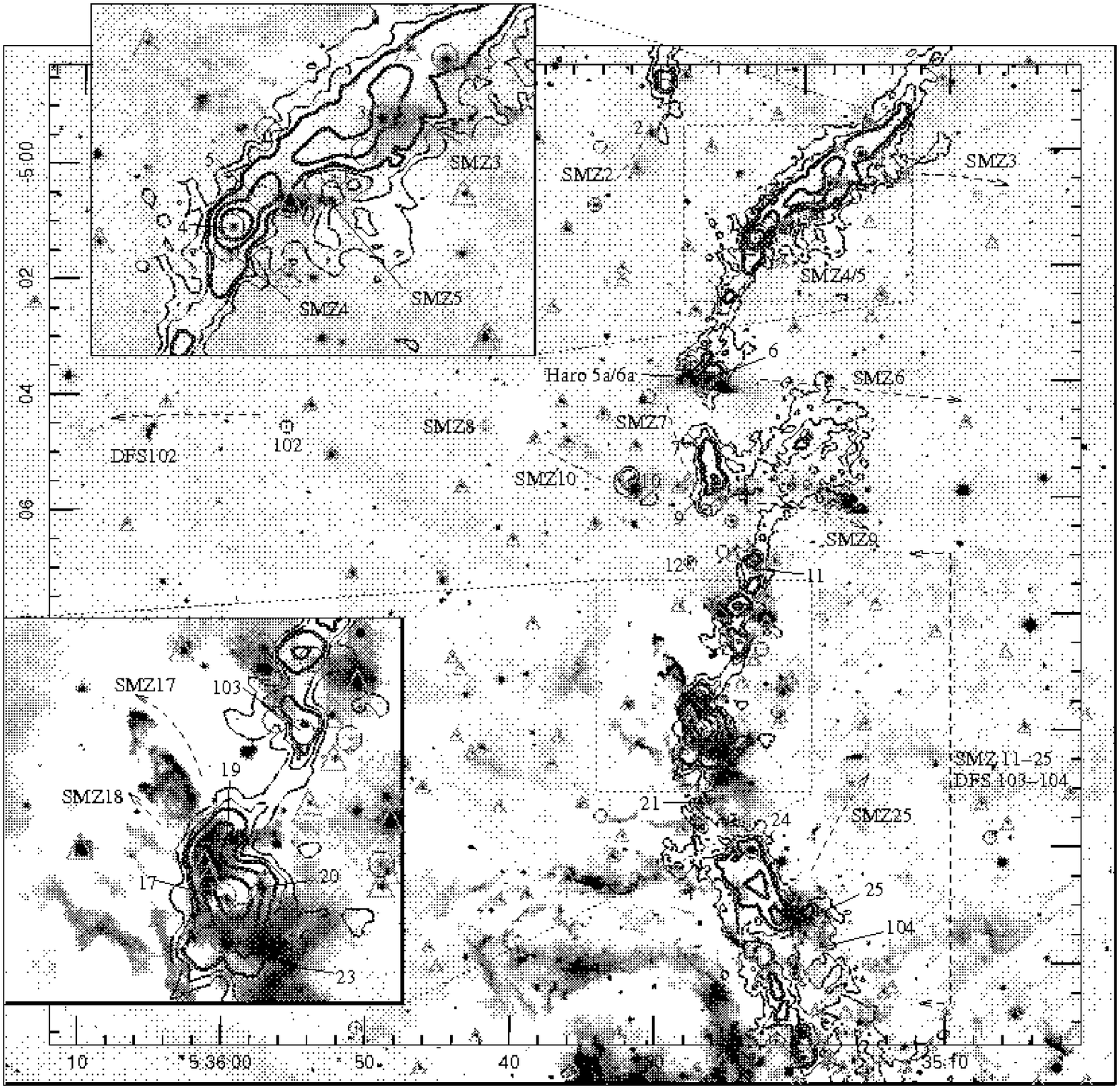}
\caption {Narrow-band \htwo\ images of OMC~2/3, the region directly
north of the ONC in Fig.~\ref{over} (inset figures are double the
scale and displayed with a logarithmic stretch).  Here and elsewhere in
this paper Spitzer protostars and disk sources are marked with circles
and triangles, respectively. MAMBO 1200\mic\ contours are
over-plotted; levels measure 5$\sigma$, 7.5$\sigma$, 10$\sigma$,
20$\sigma$, 40$\sigma$, 80$\sigma$, etc., (1$\sigma \sim 20$~mJy/beam
in central regions of the map where there is no emission). H$_2$ flows
are labelled and marked with dashed arrows (see also Sta02).}
\label{scu1}
\end{figure*}

%%%doublecolumn figure
\begin{figure*}
\centering
  \epsfxsize=17.0cm
  \epsfbox{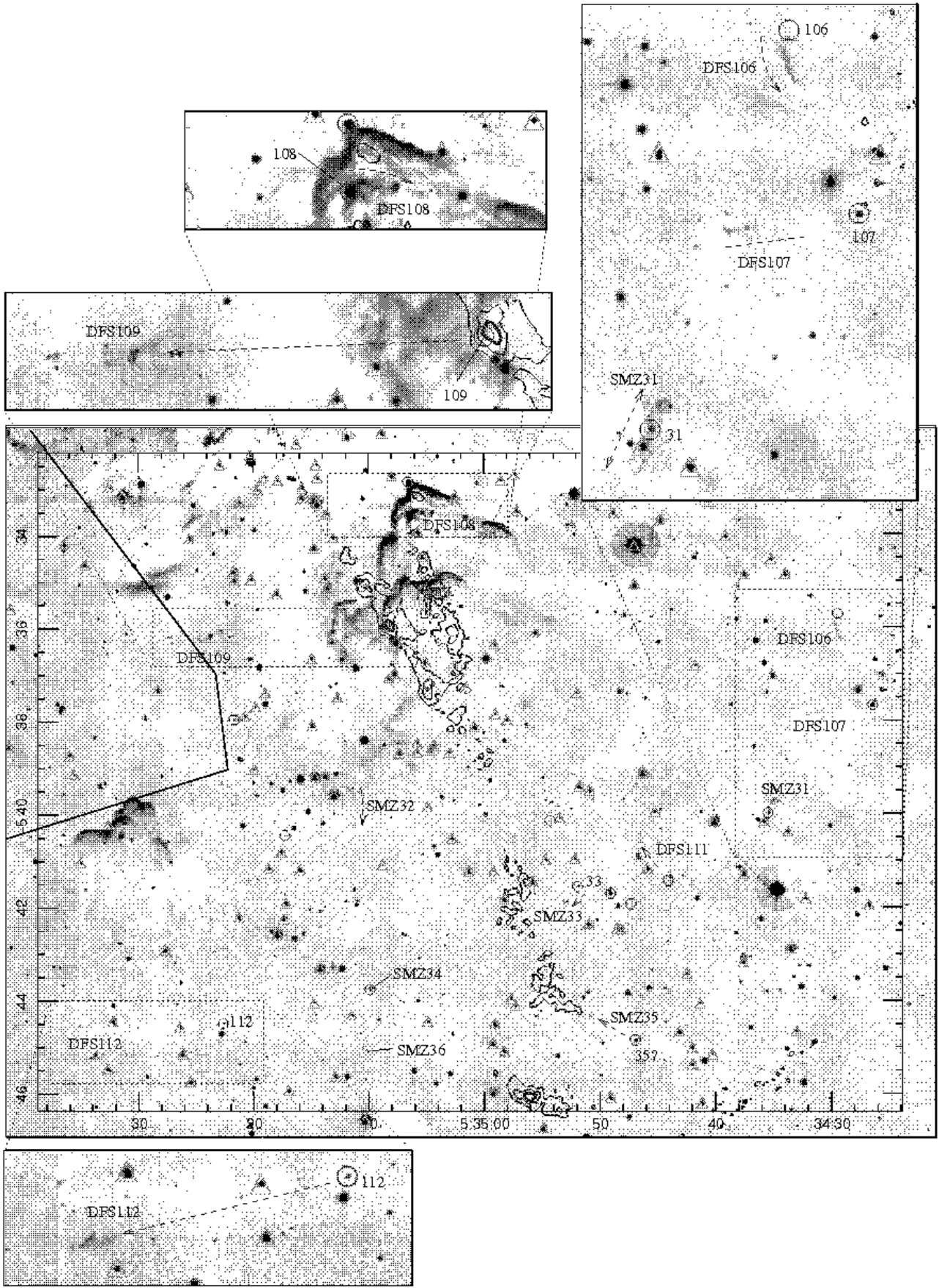}
\caption {Narrow-band \htwo\ images of the region $\sim$20\arcmin\
south of the ONC (\cite{joh06} (2006) refer to this region as OMC~4). 
Inset figures are again double the
scale and displayed with a logarithmic stretch.  
The full line on the left marks the edge of the MAMBO
map. DFS~105, DFS~110 and DFS~114 lie to the west of the region shown above.
Contours and symbols are as in Fig.~\ref{scu1}. }
\label{scu2}
\end{figure*}

%%%doublecolumn figure
\begin{figure*}
\centering
  \epsfxsize=17.0cm
  \epsfbox{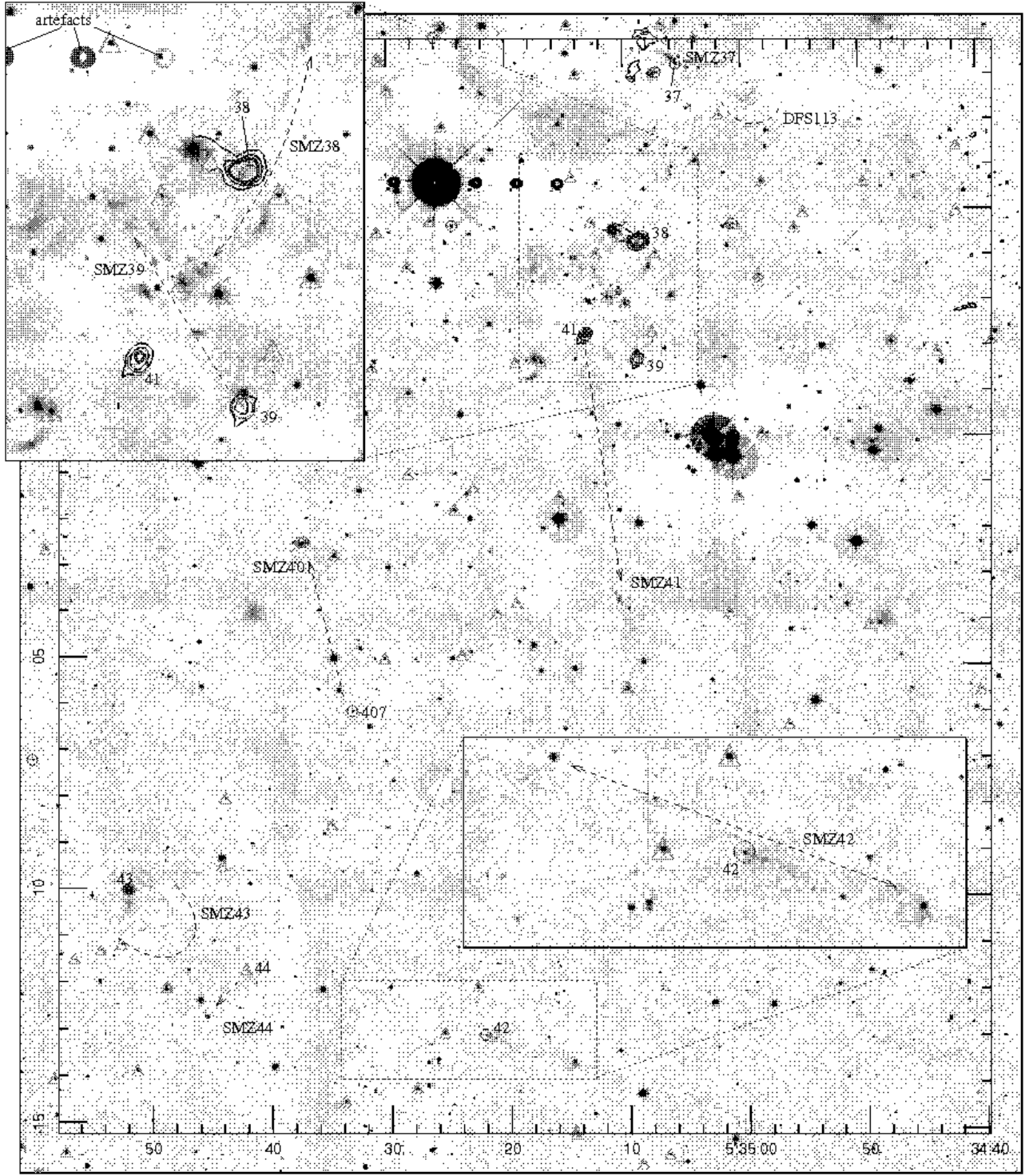}
\caption {Narrow-band \htwo\ images of the region around NGC~1980 --
OMC~5 in \cite{joh06} (2006) -- about 45\arcmin\ south of the
ONC (see Fig.~\ref{over}). Inset figures are again double the
scale and displayed with a logarithmic stretch. 
Contours and symbols are as in Fig.~\ref{scu1}.}
\label{scu3}
\end{figure*}

%%%doublecolumn figure
\begin{figure*}
\centering
  \epsfxsize=17.0cm
  \epsfbox{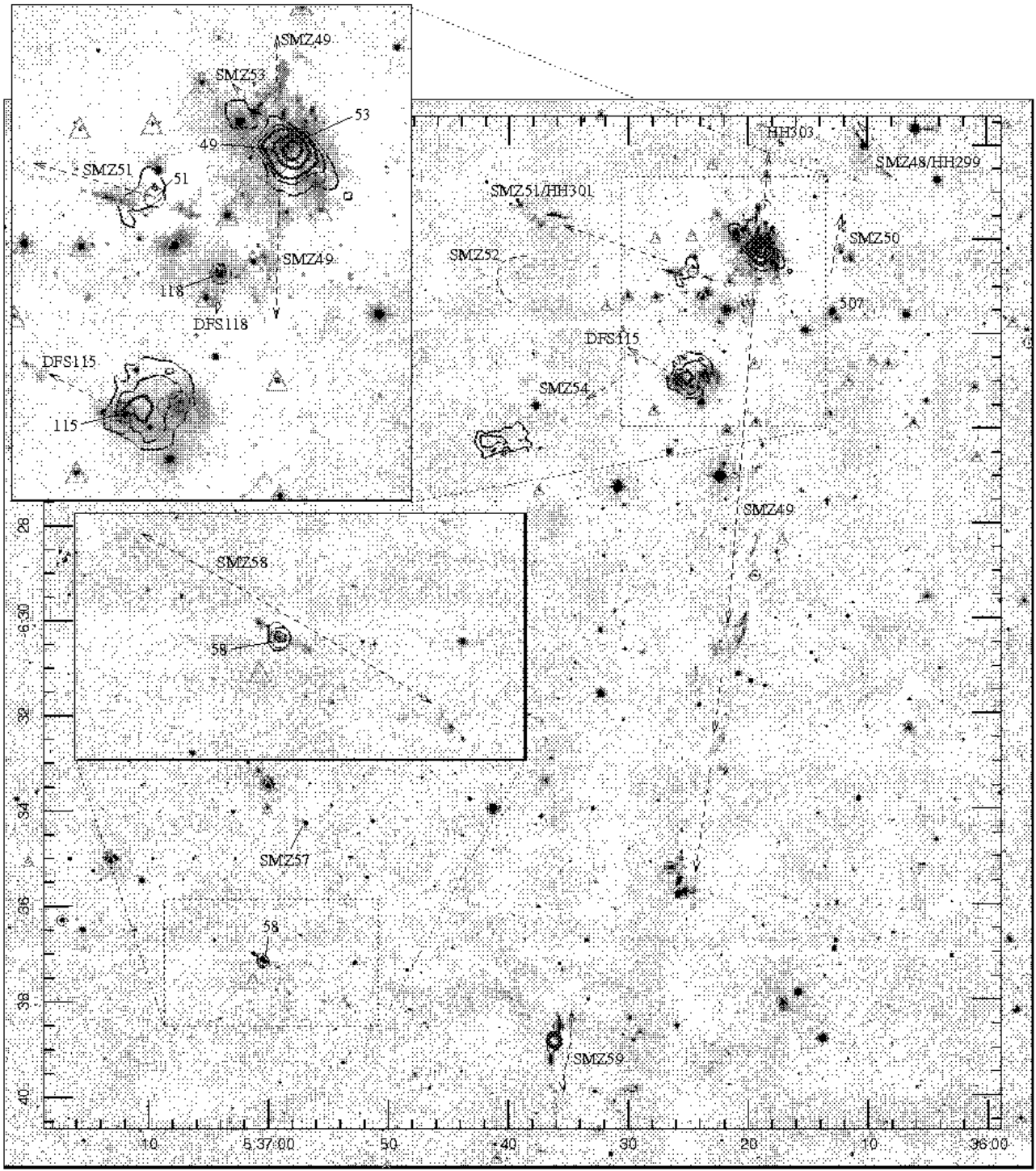}
\caption {Narrow-band \htwo\ images of L~1641-N (labelled in
Fig.~\ref{over}).  Both insets are
displayed at twice the scale and with a logarithmic stretch. 
Contours and symbols are as in Fig.~\ref{scu1}.} 
\label{scu4}
\end{figure*}

%%%doublecolumn figure
\begin{figure*}
\centering
  \epsfxsize=17.0cm
  \epsfbox{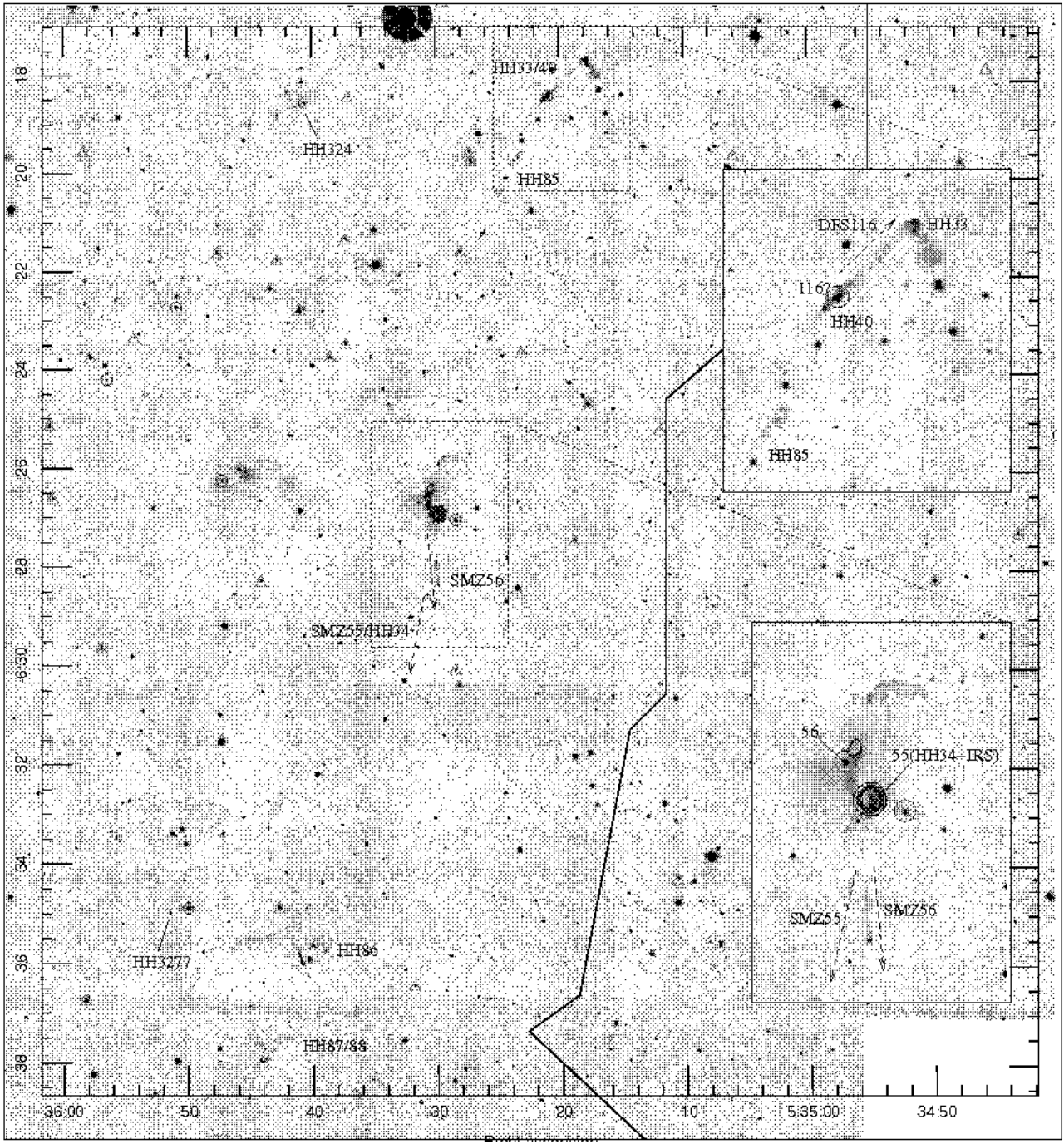}
\caption {Narrow-band \htwo\ images of the region around HH~33/40 and
HH~34 (roughly 0.5\dg\ west of L~1641-N -- see Fig.~\ref{over}).
The zig-zag line through the centre of the image marks the
western border of the MAMBO map. Both insets are again
displayed at twice the scale and with a logarithmic stretch.
Contours and symbols are as in Fig.~\ref{scu1}.} 
\label{scu5}
\end{figure*}

%%%doublecolumn figure
\begin{figure*}
\centering
  \epsfxsize=17.0cm
  \epsfbox{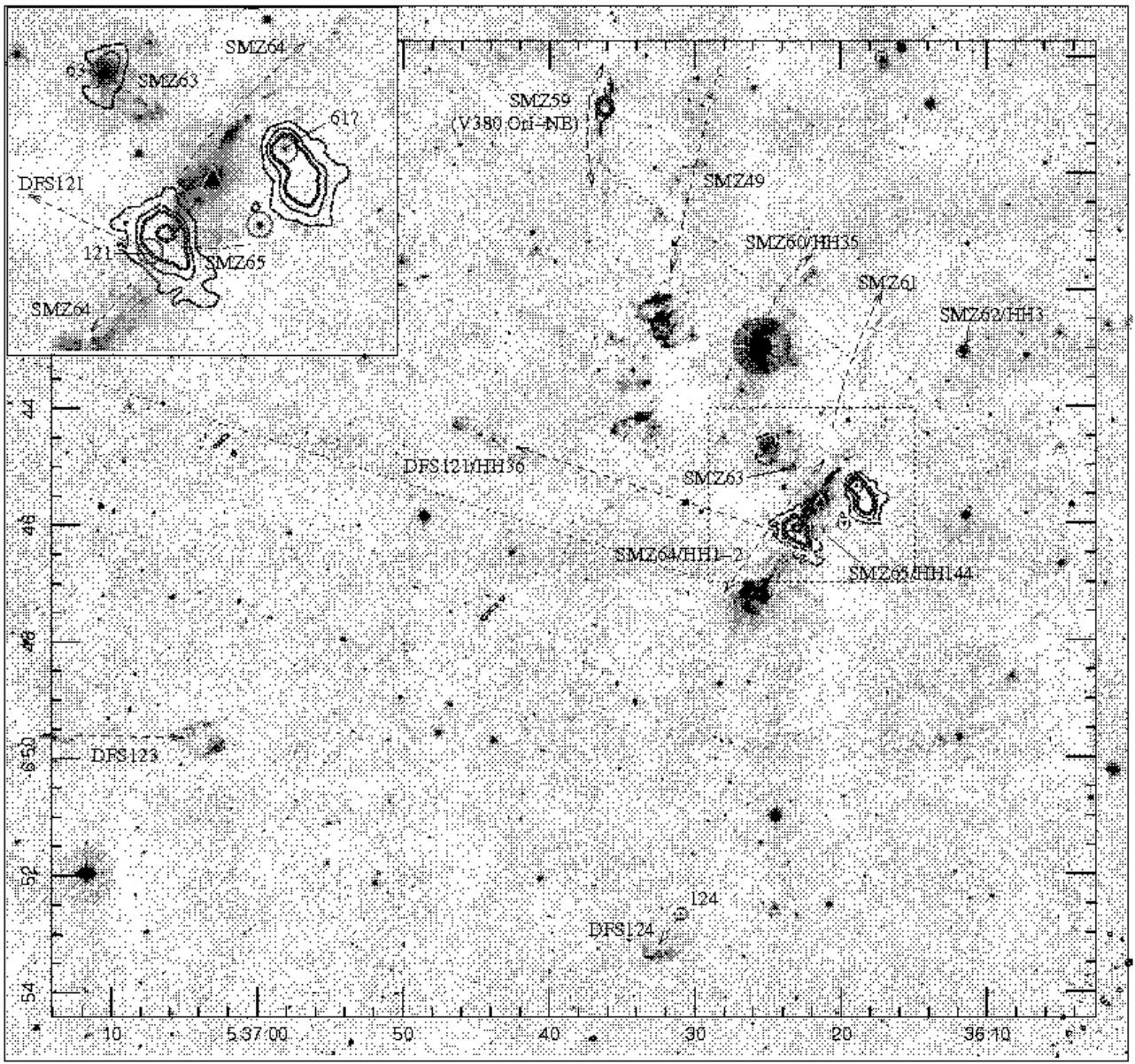}
\caption {Narrow-band \htwo\ images of the region around HH~1/2 in
NGC~1999 (roughly 0.5\dg\ south of L~1641-N -- see
Fig.~\ref{over}). The inset of HH~1 is double-scale and displayed with a
logarithmic stretch.  Contours and symbols are as in Fig.~\ref{scu1}.}
\label{scu6}
\end{figure*}

%%%doublecolumn figure
\begin{figure*}
\centering
  \epsfxsize=17.5cm
  \epsfbox{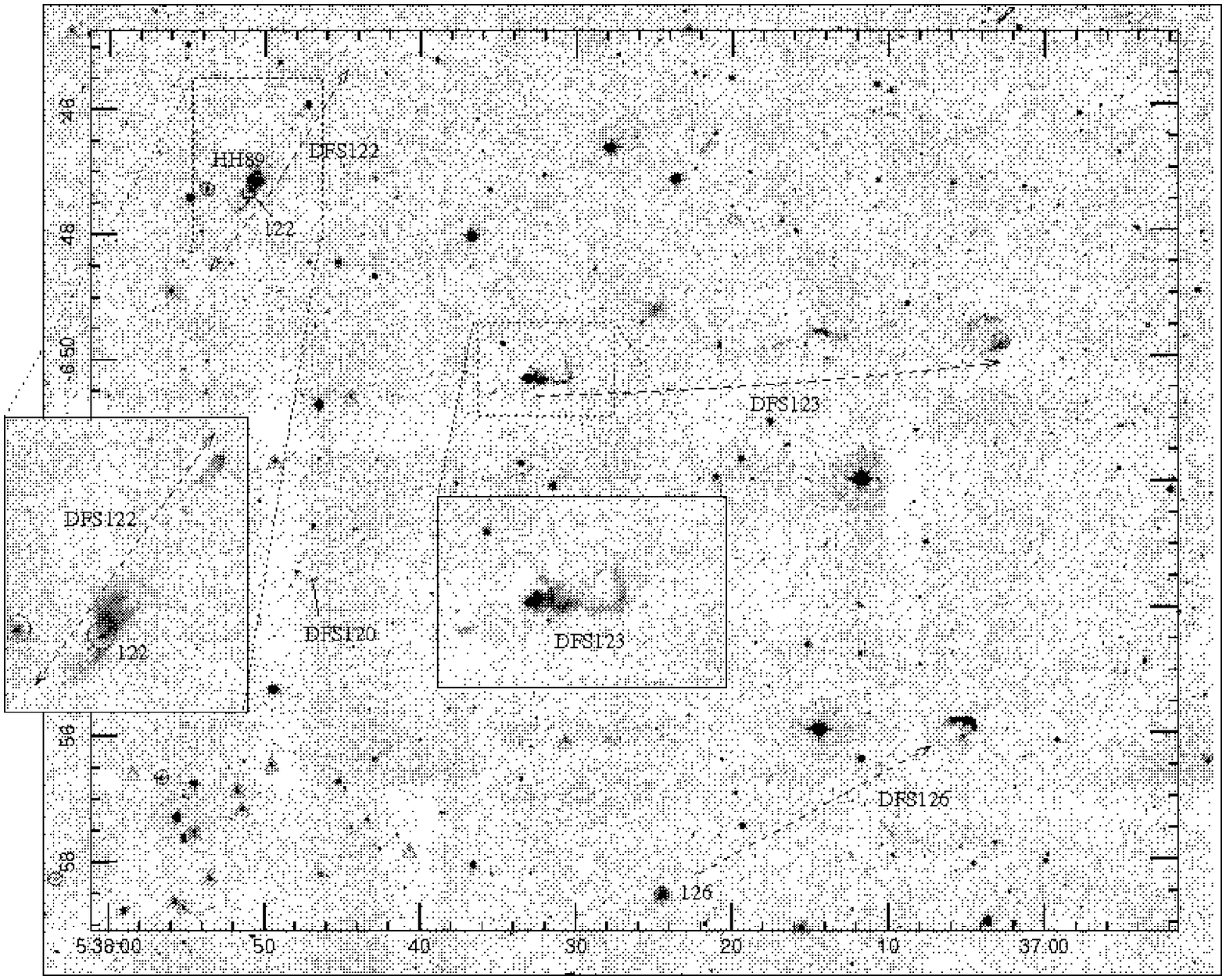}
\caption {Narrow-band \htwo\ images of the region $\sim$0.5\dg\ east of
HH~1/2 in NGC~1999 (see Fig.~\ref{over}).  Both insets are displayed
at double-scale and with a logarithmic stretch. Note that the MAMBO 
1200\mic\ (and SCUBA 850\mic ) maps do not cover this region. 
Symbols are as in Fig.~\ref{scu1}.} 
\label{scu7}
\end{figure*}

%%%doublecolumn figure
\begin{figure*}
\centering
  \epsfxsize=17.0cm
  \epsfbox{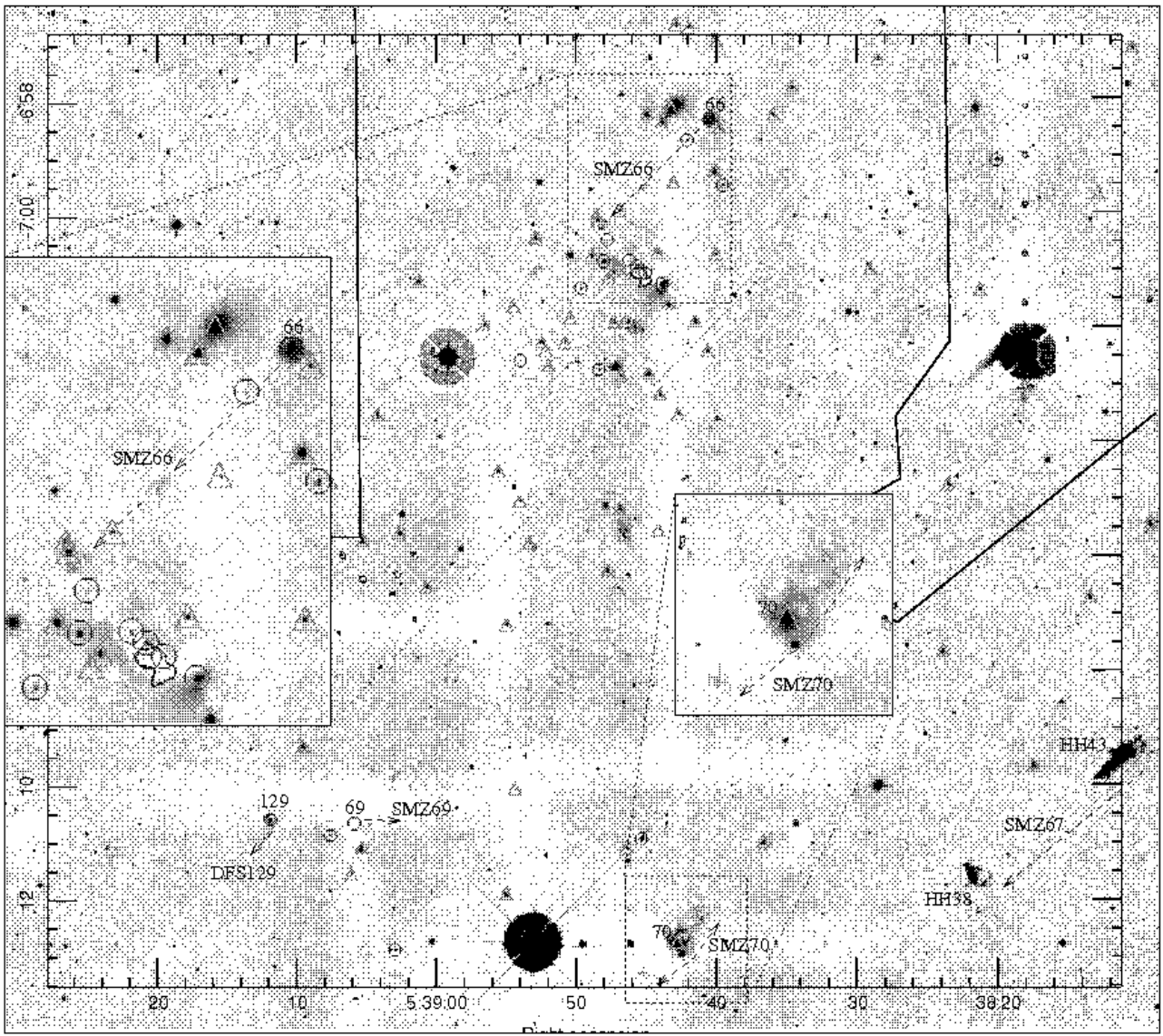}
\caption {Narrow-band \htwo\ images of L~1641-C (labelled in
Fig.~\ref{over}).  Both insets are displayed with a logarithmic
scale and at twice the scale of the main figure. 
The northeast and northwest edges of the MAMBO map are marked
with full lines. Contours and symbols are as in Fig.~\ref{scu1}.} 
\label{scu8}
\end{figure*}

%%%doublecolumn figure
\begin{figure*}
\centering
  \epsfxsize=17.0cm
  \epsfbox{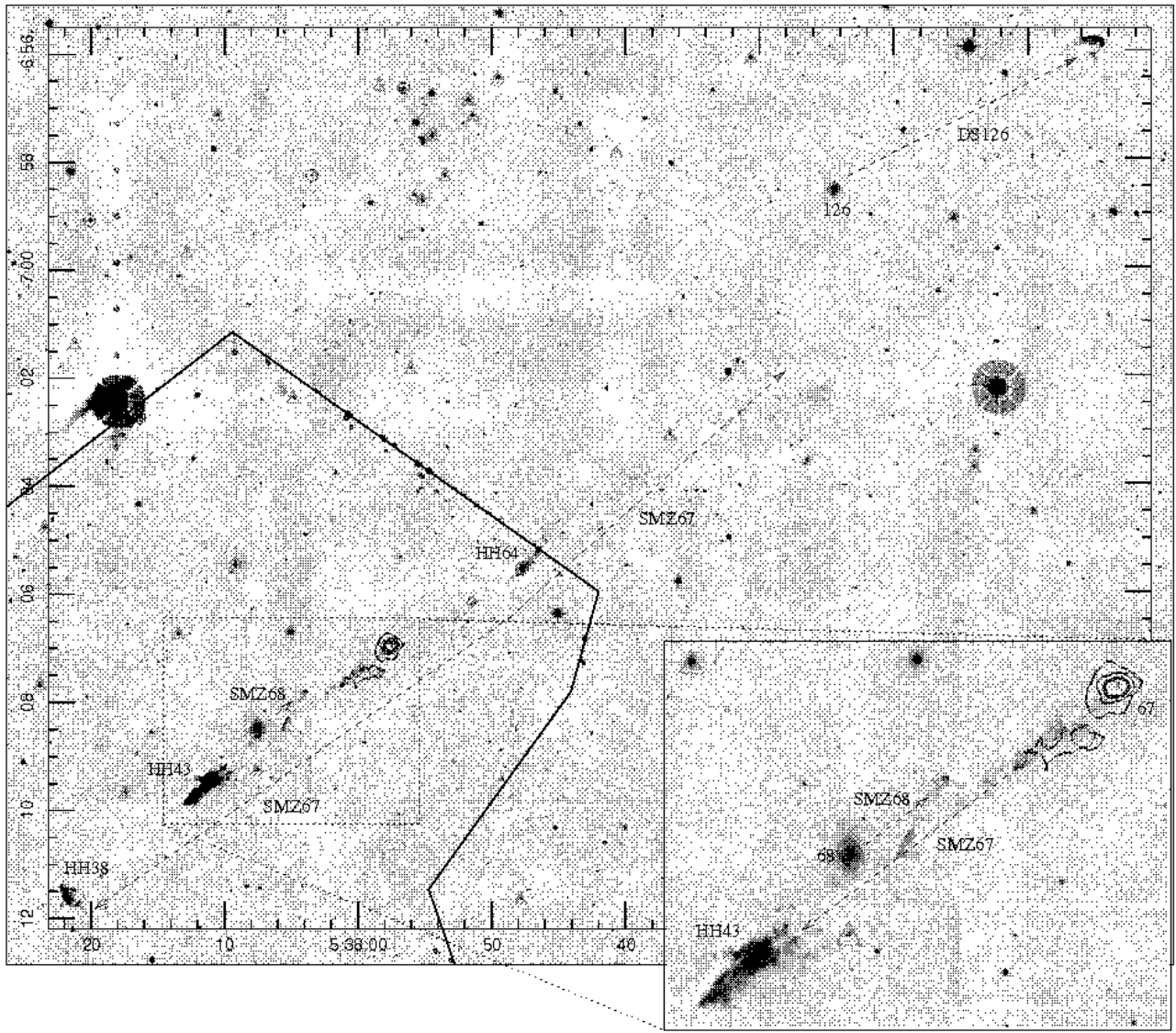}
\caption {Narrow-band \htwo\ images of the region $\sim$0.3\dg\
west-southwest of the L~1641-C cluster that includes the extensive
HH~38/43/64 bipolar outflow (see Fig.~\ref{over}).  HH~43 and its
source are shown inset at double-scale with a log stretch.  The
northern and western edges of the MAMBO map are marked with a full
line. Contours and symbols are as in Fig.~\ref{scu1}.}
\label{scu9}
\end{figure*}

%%%doublecolumn figure
\begin{figure*}
\centering
 \includegraphics[angle=90,width=155mm]
{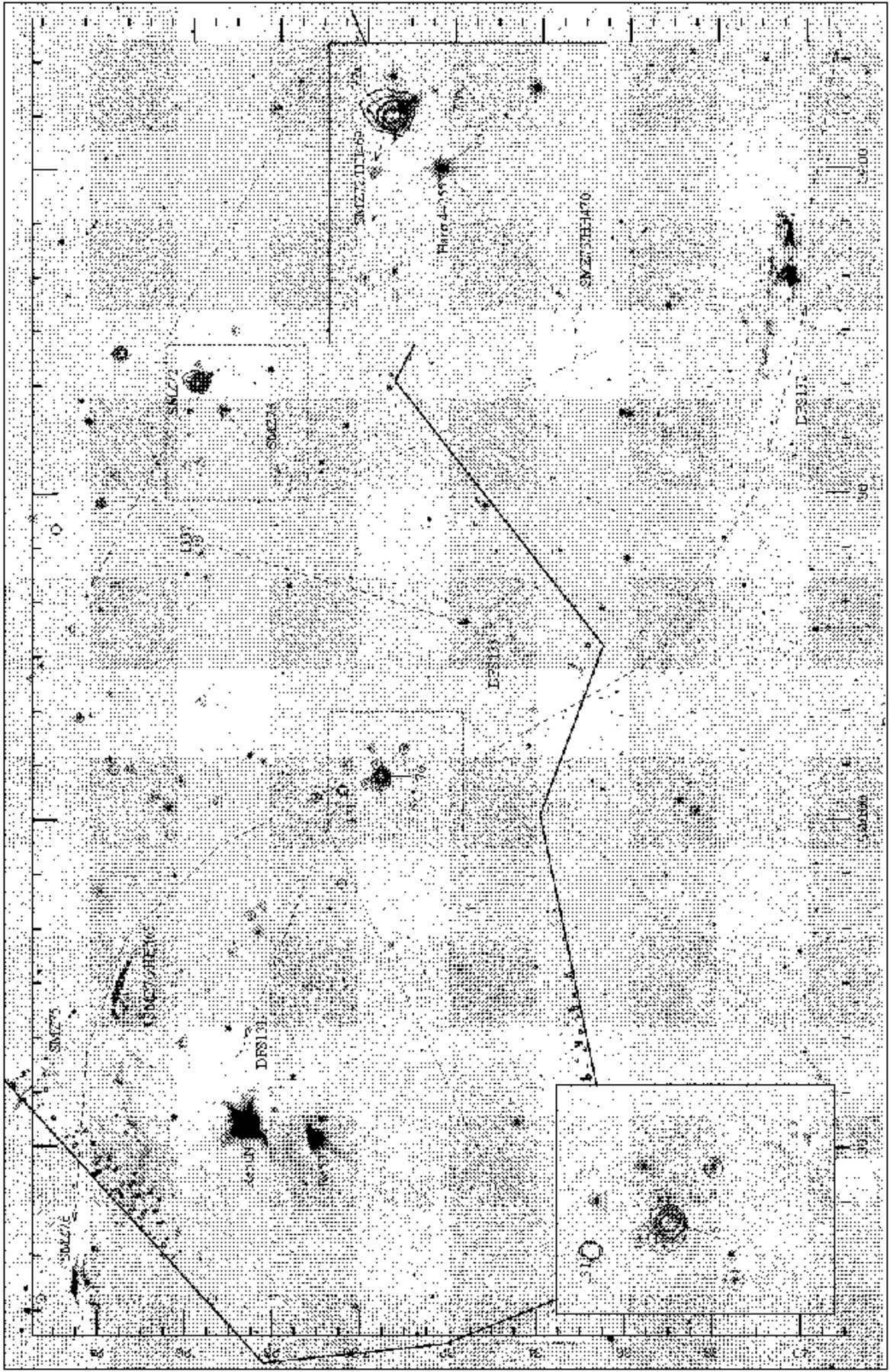}
\caption{Narrow-band \htwo\ images of the region around Re~50 and Haro
4-255, midway between L~1641-C and L~1641-S in Fig.~\ref{over}.  The
figures inset show the regions around Haro~4-255 and the likely source
of SMZ~76 (both displayed with a log stretch/double size). 
The southern edges of the MAMBO
map are marked with a full line. Contours and symbols are as 
in Fig.~\ref{scu1}.} 
\label{scu10}
\end{figure*}

%%%doublecolumn figure
\begin{figure*}
\centering
\includegraphics[angle=0,width=180mm]
{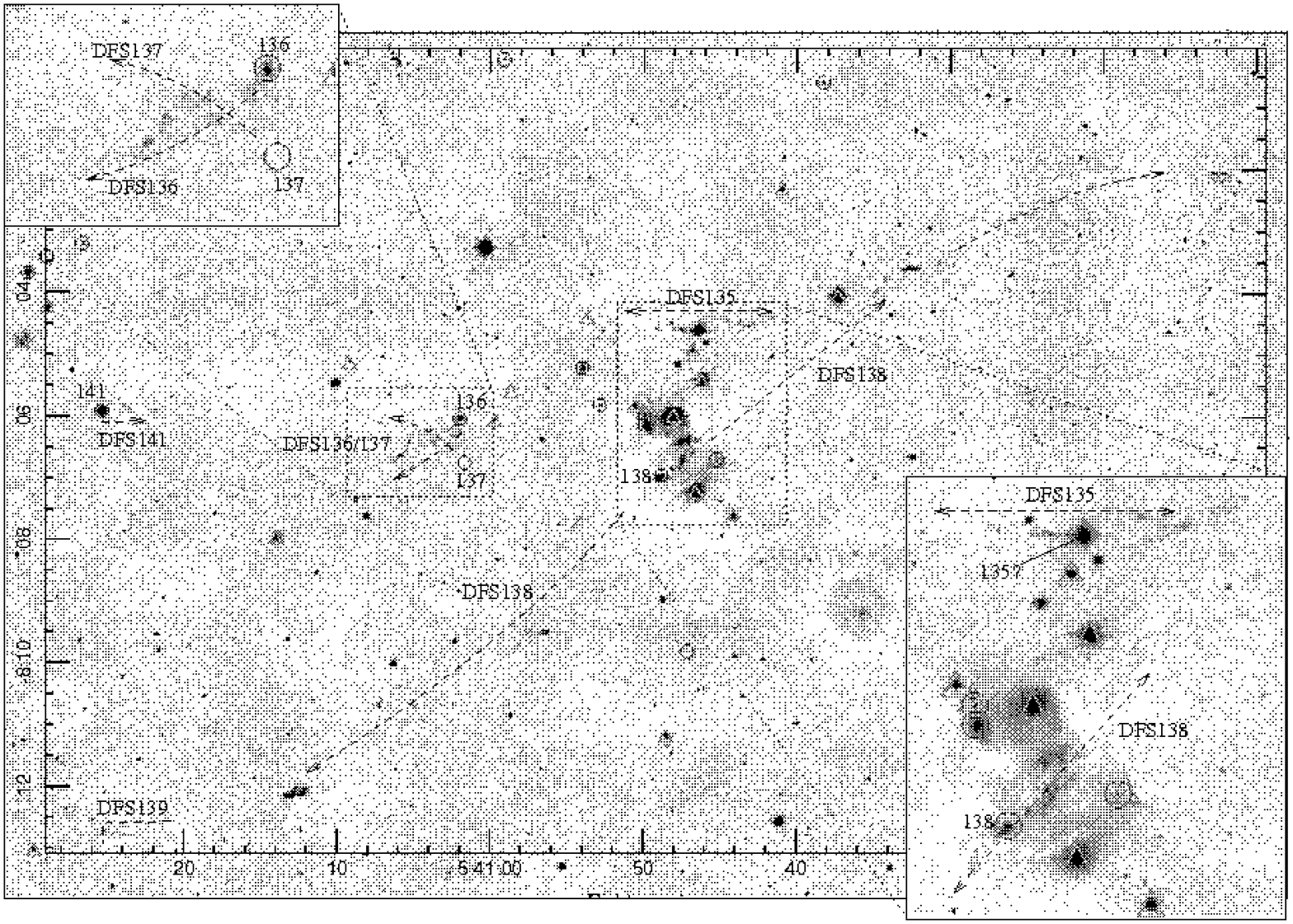}
\caption{Narrow-band \htwo\ images of L~1641-S, at the bottom of the
overview mosaics in Fig.~\ref{over}.  No MAMBO (or SCUBA) observations
are available for this region. The figures shown inset are displayed
at twice the scale and with a log stretch. Symbols are as in
Fig.~\ref{scu1}.}
\label{scu11}
\end{figure*}

%%%%%%%%%%%%%%%%%%%%%%%%%%%%%%%%%%%%%%%%%%%%%%%%%%%%%

The near-IR wide-field camera WFCAM (\cite{cas07} 2007) at UKIRT was
used on each occasion. WFCAM houses four Rockwell Hawaii-II (HgCdTe
2048x2048) arrays spaced by 94\% in the focal plane.  The pixel scale
measures 0.40\arcsec . To observe a contiguous square field on the sky
covering 0.75 square degrees -- a WFCAM ``tile'' -- observations at
four positions are required.  At each position, to correct for bad
pixels and array artifacts, a five-point jitter pattern was executed
(with offsets of 3.2\arcsec\ or 6.4\arcsec ); to fully sample the
seeing, at each jitter position a 2$\times$2 micro-stepped pattern was
also used, with the array shifted by odd-integer multiples of half a
pixel.  20 frames were thus obtained at each of the four positions in
the tile.  Eleven tiles in total were observed covering over 8 square
degrees, through both broad-band K and narrow-band H$_2$ 1-0S(1)
filters. Exposure times of 5\,sec $\times$ 2-coadds and
20\,sec $\times$ 2-coadds were used with the K and H$_2$ filters,
respectively.  With the jitter pattern and micro-stepping the total
on-source/per-pixel integration time in K was therefore 200~sec; in
H$_2$ the total exposure time was 800~sec.

The DDT data were reduced by the Cambridge Astronomical Survey Unit
(CASU), which is responsible for data processing prior to archiving
and distribution by the Wide Field Astronomy Unit (WFAU).  However,
the processing of frames that included the bright Orion Nebula and
Trapezium cluster gave images with severe background structure,
probably related to the adopted method for sky-subtraction. Our
service observations of this central region were therefore reduced
using the ORAC-DR pipeline at the telescope (\cite{cav03} 2003), which
performs flat-fielding using twilight flats, but does not subtract a
sky frame. The CASU reduction steps are described in detail by
\cite{dye06} (2006).  Residual sky/background 
structure was later removed by fitting a coarse surface to each image
(as described by \cite{dav07} 2007).  For both the CASU and Service
data, astrometric and photometric calibrations were achieved using
2MASS (\cite{dye06} 2006; \cite{hew06} 2006); the calibrated images
were subsequently used to construct the large-scale mosaics presented
in this paper.

In our K-band images we reach a limiting magnitude of $\sim$18.3 in
the less-nebulous regions; our sensitivity to point sources in the
very bright Orion nebula regions is 2--3 magnitudes worse.  In the
H$_2$ images of L~1641, outflow features with a surface brightness of
$\sim 7\times10^{-19}$~W m${-2}$ arcsec$^{-2}$ are detected at
3-5$\sigma$ above the surrounding background.  The H$_2$ sensitivity
in the ONC region is again lower because of the variable, diffuse
nebulosity.

An overview of the region observed with WFCAM is given in
Fig.~\ref{over}; colour images of the spectacular OMC~1/2/3 regions are
presented in Fig.~\ref{color}; large-scale H$_2$ 1-0S(1) 
mosaics of regions of interest throughout Orion~A are presented in
Figs.~\ref{scu1}--\ref{scu11}.  Continuum-subtracted images of the
newly-identified H$_2$ flows are available in Appendix A.  In all of these
figures axes are labelled in 2000.0 coordinates.

\subsection{Proper motion measurements}

Using the newly-acquired WFCAM images and the H$_2$ mosaics of Sta02,
which were obtained between December 1996 and May 1998 with the 3.6-m
telescope at the Calar Alto Observatory, Spain, we have measured
tangential velocities for emission-line features in a number of
flows. We focus on regions where multiple flows are observed,
particularly where the association between outflows and embedded
protostars is ambiguous. A complete list of proper motions (PMs) is
given in Appendix B, where the technique is described in detail.

\subsection{Mid-IR observations}

The Spitzer Space Telescope observations discussed in this paper were
obtained with the IRAC and MIPS cameras (\cite{faz04} 2004;
\cite{rie04} 2004).  A full description of the data analysis and
young stellar object identification will be presented in Megeath et
al. (in prep); we provide here a brief summary.  

Images in IRAC bands 1, 2, 3 and 4 (at 3.6, 4.5, 5.8 and 8.0\mic) and
MIPS band 1 (at 24\mic ) were used to compile a list of more than 300
candidate protostars between declinations -4.6\dg\ and -9.0\dg.  The
protostars were selected from their [3.6]-[4.5] and
[4.5]-[24] colours (\cite{meg09} 2009), although note that the
majority of the sources would also be classified as flat spectrum,
Class 0 and Class I protostars on the basis of their Spectral Energy
Distributions (SEDs).  To minimize contamination from extragalactic
sources we rejected all sources with protostellar colours but 24~$\mu$m
magnitudes fainter than
7.  About 20\% of the protostars were not detected at 24~$\mu$m and were
therefore identified through their [4.5]-[5.8] colour using the approach of
\cite{gut08} (2008); most of these sources are found in the Orion
nebula where the 24~$\mu$m data were saturated.

There is a small amount of contamination from extragalactic sources,
particularly AGN; we estimate this contamination to be 0.7 sources per
sq. degree. In addition, there may also be some contamination from
embedded Class II sources and potentially from edge-on young stars
with flared disks.  Estimates of the degree of contamination await the
analysis of recent Spitzer 5-40~$\mu$m spectroscopy of the sample. In
addition to the protostars, a further $\sim$ 2000 disk excess sources
were identified in the Spitzer data.

The sample of protostars is not
complete: we expect that many protostars are hidden in the bright
nebulosity toward the Orion nebula, and that several very bright
prototstars were omitted because of strongly saturation in the
24~$\mu$m data.

To characterize the SED of each protostar and disk excess source, a
spectral index was obtained by fitting a single power-law to the 3.6,
4.5, 5.8, 8 and 24~$\mu$m photometry. For sources without 24~$\mu$m
detections, the spectral index was calculated from only the IRAC
photometry. The resulting 3.6-24~$\mu$m spectral-index, $\alpha = d
log (\lambda F(\lambda))/d log \lambda$, ranges from -0.5 to 3.0 for
the protostars.  Although the spectral index is useful as an indicator
of evolution, the index does depend on the density, the rotation rate,
and the inclination of the protostellar envelope (\cite{whi03}
2003,2004).  Consequently, one should not expect a fool-proof
relationship between $\alpha$ and the age of the protostar. A
histogram showing the distribution of $\alpha$ for the protostars in
Orion peaks at around 0 and gradually declines with increasing
$\alpha$.  In contrast, the disk excess sources have spectral indices
in the range $-2.5 < \alpha < 0.0$; a histogram of $\alpha$ values for
these objects peaks at around -0.8, consistent with a large population
of Class II objects.  We note that the spectral indices of the disk
excess sources can be affected by extinction.  Hence, the distribution
of their $\alpha$ values overlaps the distribution of $\alpha$ for the
protostars (note that there is not a discontinuity in the distribution
of $\alpha$ between the disk excess and protostars).  This may reflect
a continuous transition between protostars and disk excess sources, as
the infalling envelope is dissipated and the inner star/disk system is
revealed.

\subsection{(Sub)millimetre observations}

In many of the subsequent figures we also overlay contours of dust
continuum emission at 1200\mic .  These data were obtained with the
IRAM 30-m telescope at Pico Veleta, Spain, with the 37 and 117-pixel
Max Planck Millimeter Bolometer (MAMBO) arrays (\cite{kre99}
1999). The data were acquired over several observing runs between 1999
and 2002, and reduced using the in-house reduction package ``MOPSI''.
Standard reduction steps were employed; baseline-fitting, de-spiking,
correction for atmospheric extinction and flux calibration using
planetary observations.  At IRAM the MAMBO Half Power Beam Width
(HPBW) measures 11\arcsec .

A preliminary catalogue of $\sim$500 MAMBO cores was obtained from visual
inspection of the map and 2-dimensional Gaussian fitting of features.
From this list we identify cores that coincide with an H$_2$
outflow source and list the core parameters in Tables~\ref{smz} and
\ref{dfs}. The MAMBO dataset will be analysed in detail in a future paper
(Stanke et al., in prep.).

%%%%%%%%%%%%%%%%%%%%%%%%%%%%%%%%%%%%%%%%%%%%%%%%%%%%%%%%%%%%%
%%%%%%%%%%%%%%%%%%%%%%%%%%%%%%%%%%%%%%%%%%%%%%%%%%%%%%%%%%%%%

\section{Results}

\subsection{Overview}

In Fig.~\ref{over} we present a mosaic of the complete K-band dataset.
We indicate on this figure the extent of the 850\mic\ and 1200\mic\
maps, and mark the positions of $\sim$300 protostars.

Colour images of the bright, nebulous regions around the ONC/M~42, the
M~43 HII region, and the complex star forming environment around Haro
5a/6a are shown in Fig.~\ref{color}.  These allow the reader to better
distinguish H$_2$-emitting jets and outflows from filaments of bright
nebulosity.  Near-IR observations of the region around BN/KL,
including the spectacular Orion Bullets, the Orion Bar, BN and the
Trapezium cluster, have been discussed in many other papers
(e.g. \cite{usu96} 1996; \cite{sch97} 1997; \cite{lee00} 2000;
\cite{bal00} 2000; \cite{car01} 2001; \cite{sle04} 2004; \cite{kas06} 2006; 
\cite{sim06} 2006; \cite{tam06} 2006).  
In this paper we therefore focus on the area to the north of
the ONC, the OMC~2/3 region, and the extensive L~1641 low-mass star
forming region to the south.

In Figs.~\ref{scu1}-\ref{scu11} we present segments of the full H$_2$
mosaic, indicating with dashed lines the locations of H$_2$ jets and
outflows (the data in these figures have been binned over
4$\times$4 pixels, to a pixel scale of 0.8\arcsec , to enhance the fainter
features).  Many of the outflows imaged here have already been
catalogued by Sta02 (H$_2$ images of OMC~2/3 are also presented
by \cite{yu00} 1997 and \cite{yu00} 2000).  With the newly-identified H$_2$ 
flows, we therefore continue the numbering scheme of Sta02, although
new flows (often found beyond the bounds of the Sta02 data) are
numbered above 100.  The Sta02 objects and the new H$_2$ outflows are
prefixed with SMZ and DFS, respectively. Note that entire H$_2$ flows,
rather than individual H$_2$ features or small groups of knots, are
given a single number.  For example, HH~1 and HH~2, which are
relatively discrete objects, are here (and in Sta02) referred to as
SMZ~64. All H$_2$ flows are listed in Tables~\ref{smz} and \ref{dfs};
continuum-subtracted H$_2$ images of the new DFS flows are given in
Appendix A, along with a brief description of each outflow.

In Figs.~\ref{scu1}-\ref{scu11} we also mark the positions of the
protostars and (more numerous) disk excess sources identified from the
Spitzer data; candidate outflow sources, when identified, are numbered
using the same value as is given to the H$_2$ outflow itself (so for
example DFS~101 is driven by IRS~101). Outflow sources are listed in
Tables~\ref{smz} and \ref{dfs}, together with any coincident molecular
cores seen at 850\mic\ or 1200\mic .  850\mic\ cores are taken from
\cite{nut07} (2007); the table of 1200\mic\ cores has yet to be
published (Stanke et al., in prep.), so here we again use the outflow
number, together with MMS (for MilliMetre Source), to identify each
core.

The association of jets with young stars and/or (sub)millimetre
continuum peaks is largely based on morphology and/or
alignment. However, in some of the more complex regions we use the PM
measurements in Appendix B to help isolate the location of the outflow
source; in most cases the direction of propagation of the shock
features are as one might expect, based on their morphology and the
location of candidate outflow sources.

%%%%%%%%%%%%
%%%%%%%%%%%%

\subsection{Outflow activity in Orion~A}

OMC~2/3 (Figs.~\ref{color} and \ref{scu1}), the region to the north of
the ONC, is described in considerable detail by \cite{yu00} (2000).
HH objects are identified by \cite{rei97} (1997) and \cite{bal01}
(2001).  High-resolution CO outflow maps are presented by
\cite{wil03} (2003).  The distribution of 
dense molecular gas throughout OMC~1/2/3 has been mapped at 450\mic\
and 850\mic\ by \cite{joh99} (1999);
\cite{chi97} (1997) present 1300\mic\ observations of OMC~2/3.
A comprehensive review is also given by \cite{pet08} (2008).

The vast majority of protostars in Fig.~\ref{scu1} are located within
11\arcsec\ of a MAMBO core and/or 14\arcsec\ of a SCUBA core (these
radii being equivalent to the HPBW of these
(sub)millimetre data).  The association of bright H$_2$ jets with
Spitzer protostars and cold molecular cores attests to the youth of
the outflows in this region.  Most of the protostars are located along
the chain of cores that runs through the centre of OMC~2/3 (the disk
excess sources are more widely distributed); similarly, most of the
H$_2$ outflows seem to be driven by sources close to this central
axis.  The most distinctive H$_2$ flows, SMZ~3, SMZ~5, SMZ~6 and SMZ~9 are
also orientated perpendicular to the north-south axis of cores.

From Sta02 there are 22 H$_2$ flows spread across the region shown in
Fig.~\ref{scu1} (we regard SMZ~11 and SMZ~13 as a single flow;
likewise SMZ~14/16 and SMZ~21/22).  We identify four new H$_2$ flows north
of the ONC, DFS~101-104.  Candidate outflow sources are found for 24
of the 26 H$_2$ flows.  Only two have negative spectral indices; the rest
are either ``flat spectrum'' or reddened protostars.  Two of the 25
Spitzer outflow sources were not observed with MAMBO; of the remaining
23, 18 are associated with 1200\mic\ (MMS) cores.

In Figs.~\ref{scu2} and \ref{scu3} we show large-scale H$_2$ images of
two regions south of the ONC, adjacent to NGC~1980 in the overview
plot in Fig.~\ref{over}. \cite{joh06} (2006) refer to these
regions as OMC~4 and OMC~5, respectively. In and around these two
areas Sta02 identify at least 18 possible H$_2$ flows (SMZ~29--SMZ~46;
between declination -5\dg 20\arcmin\ and -6\dg 15\arcmin ); we add a
further ten H$_2$ flows to this tally (DFS~105--114; note that
DFS~105, DFS~110 and DFS~114 are beyond the edges of
Fig.~\ref{scu2}). Moreover, PMs have been measured for many of the
flows in this area.

We identify candidate outflow sources for 25 of the 28 outflows in the
OMC~4 and 5 regions, although unlike OMC~2/3, many of these sources are
not associated with dust cores.  (Note that \cite{joh06} (2006) also
compare H$_2$ images with dust continuum maps in this region, using
the images of Sta02 and their SCUBA observations, respectively.) All
of the SMZ jets were observed with MAMBO, as were seven of the 10 DFS
flows.  However, only 14 of the 25 observed H$_2$ outflow sources are
associated with 1200\mic\ cores. There are clearly considerable
differences between the star forming regions north of the ONC (OMC~2/3), and
those directly south (OMC~4/5).

As we move further south (through L~1641-N towards L~1641-C, see
Fig.~\ref{over}) fewer protostars are identified in the Spitzer
analysis.  Even so, many sources are clearly driving H$_2$ flows: in many
regions (e.g. around L~1641-N, HH~34 and HH~1/2 in Figs.~\ref{scu4},
\ref{scu5} and \ref{scu6}) we note a clear association between 
Spitzer protostars, (sub)millimetre cores and H$_2$ outflows.

The infrared jets and outflows in the very busy region around L~1641-N
(Fig.\ref{scu4}) are described in detail by \cite{gal07} (2007) and
\cite{sta07} (2007).  The former present Spitzer observations and PMs
for knots close to L~1641-N itself; the latter show high-resolution CO
maps that reveal a number of bipolar outflows centred on this cluster.
The HH objects in the region are discussed by \cite{rei98}
(1998). Sta02 identify seven H$_2$ flows in and around L~1641-N
(SMZ~48--SMZ~54); we label two additional flows, DFS~118 -- which
is associated with a bipolar CO outflow (\cite{sta07} 2007), and
DFS~115 -- which is associated with a small cluster of protostars
$\sim$3\arcmin\ south of L~1641-N, known as Strom 11.  Six of these
nine H$_2$ flows have candidate Spitzer protostellar sources; all but one of
these protostars is associated with 1200\mic\ emission or a MAMBO
core.  In total there are 10 protostars within a 5\arcmin\ radius of
L~1641-N; it seems likely that most, if not all of these sources could
be associated with H$_2$ line emission features.

The region to the west of L~1641-N, around HH~34, is shown in
Fig.\ref{scu5}.  The source of the well-known HH~34 jet, HH~34-IRS
(IRS~55), is flagged as a protostar in the Spitzer data, along with
two other sources, one of which (IRS~56) drives a second H$_2$ flow.
North of HH~34, the well-known HH objects HH~33/40 have in recent
years been considered to be part of a parsec-scale flow that includes
HH~34 and HH~86/87/88 to the south (\cite{dev97} 1997; \cite{eis97}
1997).  However, in the Spitzer analysis a protostar (IRS~116) is
found coincident with HH~40. The source is point-like in 24\mic\ MIPS
data so is probably not a mis-identified H$_2$
features\footnote{H$_2$ features produce pure-rotational emission
lines in each of the four Spitzer IRAC bands, though not in the MIPS
24\mic\ band (e.g. \cite{smi05} 2005; \cite{vel07} 2007).  Spectral
indices measured from the IRAC bands alone can therefore mimic values
expected for protostars (\cite{dav08} 2008), though a MIPS detection
would then require bright, forbidden line emission from [FeII].}.
IRS~116 is not associated with a dense SCUBA or MAMBO core.  Even so,
it remains a viable candidate outflow source for HH~33/40.

Roughly 25\arcmin\ west of HH~34, HH~83/84 (\cite{rei89} 1989) is also
observed in H$_2$ emission.  HH~83-IRS, the source of this well-known
jet, is outside the bounds of the SCUBA and MAMBO observations, and
the Spitzer IRAC 3.6\mic\ and 5.8\mic\ and MIPS 24\mic\ maps, although
a bright source is detected in the remaining IRAC bands with an
infrared excess. HH~83-IRS's [4.5]-[8] color is more consistent with a
disk excess source than a protostar.

The outflows in the vicinity of HH~1/2, shown here in Fig.~\ref{scu6},
are well documented (see e.g. \cite{dav94} 1994; \cite{hes98} 1998;
\cite{rei00} 2000).  Radial velocities and PMs -- based on optical
observations -- have been reported in the literature for the central
bright HH flows (\cite{eis94} 1994; \cite{bal02} 2002), so we have
only measured PMs for the bright H$_2$ features 4\arcmin --5\arcmin\
northeast of HH~1/2.  We find that these knots are moving southward,
which confirms their proposed association with the large-scale SMZ~49
outflow. Around HH~1/2 itself, only four Spitzer protostars are
identified (SMZ~62/HH~3 is probably a mis-identified H$_2$
knot).  Three of the four protostars are associated with 1200\mic\
cores; all four sources likely drive H$_2$ flows.  The well-known VLA
source of the HH~1/2 outflow is unresolved from neighbouring sources
and therefore is not listed as an ``IRS'' source in Table~\ref{smz}.

By extending the maps to the southeast and southwest of the HH~1/2
region we discover six more H$_2$ flows, DFS~120 and DFS~122--DFS~126,
four of which are shown in Fig.~\ref{scu7} (DFS~122 is associated
with HH~89; \cite{rei85} 1985).  Three of the six have candidate
protostellar sources (one was not observed with Spitzer). With a
length of 1.1~pc, DFS~123 qualifies as a ``parsec-scale'' flow.

Roughly 30\arcmin\ to the southeast of HH~1/2, the Spitzer data
recover the cluster of young stars collectively known as L~1641-C
(Fig.~\ref{scu8}). However, like Sta02 we identify only one clear-cut
flow in the region, SMZ~66.  The cluster is associated with only faint
1200~\mic\ emission and so L~1641-C probably harbours a fairly evolved
population.  The absence of ambient material explains the lack of
H$_2$ flows.

Fig.\ref{scu9} shows the parsec-scale flow HH~38/43/64 (SMZ~67).  The
source of this outflow, and the neighbouring H$_2$ jet SMZ~68, are
both retrieved from the Spitzer data.  These protostars are also
associated with molecular cores.  If we include the DFS~126 bow-shock
in the HH~38/43/64 outflow, then the total length of this system is
25\arcmin\ (3.2~pc).

The region midway between L~1641-C and L~1641-S (Fig.\ref{scu10}; see
also Fig.~\ref{over} for reference) includes the well-known T Tauri
star Haro~4-255 (\cite{asp00} 2000), its infrared neighbour
Haro~4-255 FIR, and the spectacular arc of H$_2$ emission, SMZ~76,
first noted by \cite{sta00} (2000).  SMZ~74 (not shown here), SMZ~75
and SMZ~76 all have morphologies and PMs that imply motions away from
the chain of young stars near the centre of Fig.\ref{scu10}, although
precise associations are difficult to make.  \cite{sta00} (2000)
identified a millimetre peak southwest of SMZ~76 (which he labeled
L~1641-S3~MMS1) as the location of the likely source of this object;
we note that DFS~132 may be the counter-lobe of this remarkable
outflow.  The protostar/dust core IRS~131/MMS~131 may also drive an
H$_2$ flow (DFS~131), though there are perhaps a dozen Spitzer YSOs in
this region with no obvious H$_2$ jet.  Most have no associated dust
core and are therefore, like the young stars in L~1641-C, probably too
evolved.

Lastly, the region at the southern end of L~1641 is abundant with
Spitzer-identified protostars and newly discovered H$_2$ flows. In
Fig.\ref{scu11} we label six possible outflows, four of which have
candidate sources (note that the MAMBO observations do not extend this
far south in L~1641).  DFS~138 is the most spectacular H$_2$ flow,
comprising a bright, collimated, bipolar jet and sweeping bow shocks.
The flow extends over 13.4\arcmin\ (1.8~pc) and is probably driven by
the Spitzer protostar labelled IRS~138.

%%%%%%%%
%%%%%%%%

 \begin{figure}
   \centering
   \includegraphics[width=8.5cm]
            {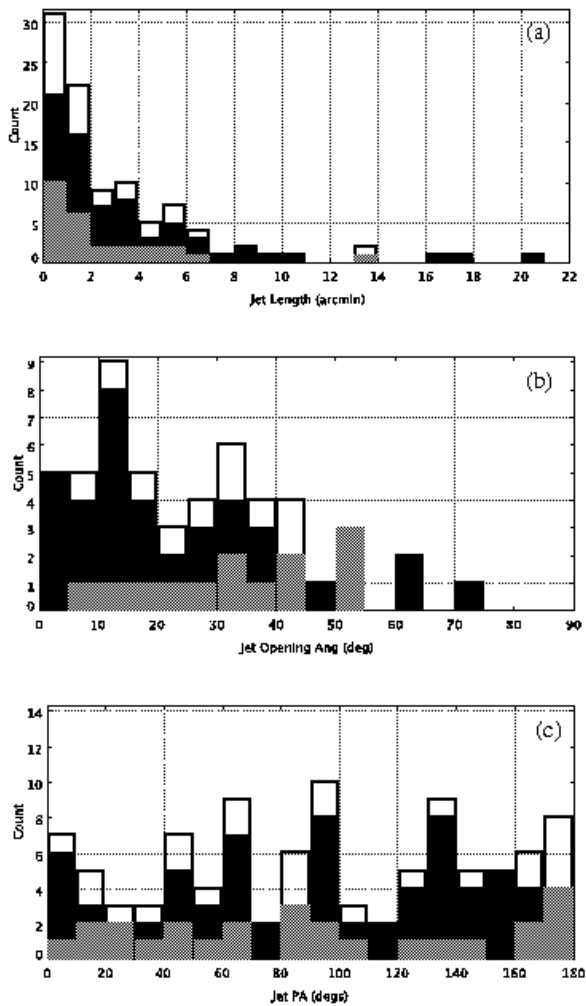}
    \caption {Histograms showing (a) the distribution of jet lengths (in
      1\arcmin\ bins),
      (b) jet opening angles (5\dg\ bins), and (c) jet position angles
          (10\dg\ bins), for all H$_2$ jets in Orion~A (open columns),
          for H$_2$ flows north of
      the ONC (declination $>$ -5\dg 30\arcmin ; grey filled columns)
      and for H$_2$ flows south of the ONC ($<$ -5\dg 30\arcmin ; black
      filled columns). }
    \label{histograms}
 \end{figure}

 \begin{figure}
   \centering
   \includegraphics[width=7.5cm]
            {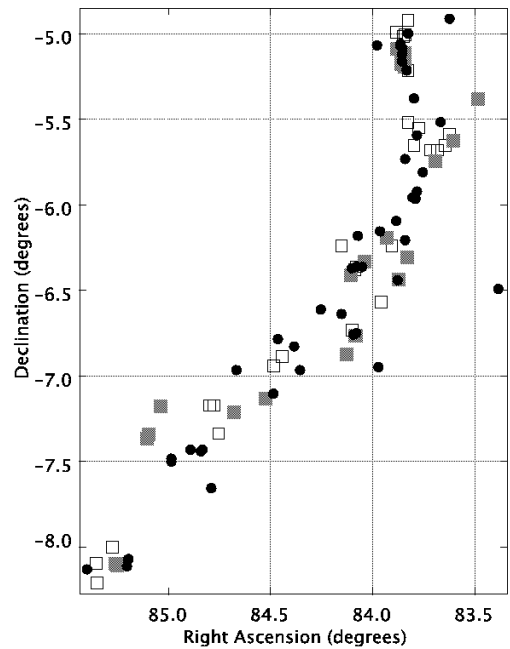}
    \caption {A map showing the distribution of short ($L <$1\arcmin ;
    open squares),
              intermediate (1\arcmin $< L <$ 2\arcmin ; filled grey
              squares), and long ($L >$2\arcmin ; filled black
              circles) H$_2$ jets.}
    \label{lenmap}
 \end{figure}

 \begin{figure*}
   \centering
   \includegraphics[width=17cm]
           {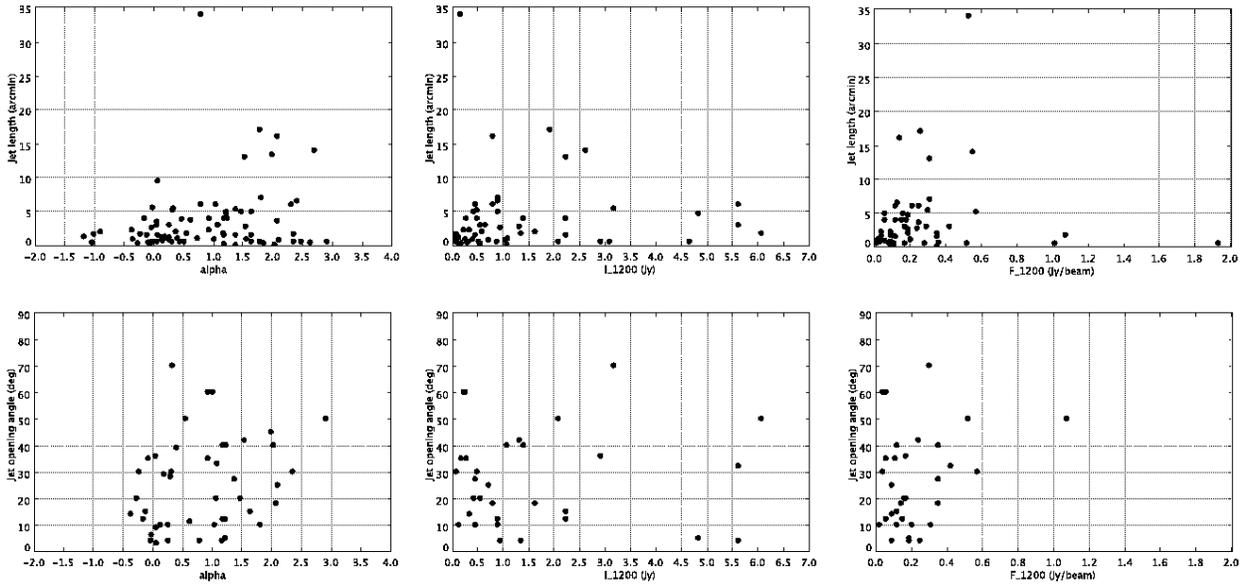}
    \caption {Source spectral index, $\alpha$, integrated 1200\mic\ core 
       flux, $I_{1200}$, and 1200\mic\ flux density, $F_{1200}$,
       plotted against H$_2$ jet length and
       jet opening angle for all jets in Orion A with identified sources
      (Tables~\ref{smz} and \ref{dfs}). }
    \label{params}
 \end{figure*}

%%%%%%%%
%%%%%%%%

\subsection{Outflow statistics}

In Tables~\ref{smz} and \ref{dfs} we list 116 jets, of which 43 are
newly identified.  Of the 111 flows that were within the bounds of the
Spitzer observations, 72 (62\%) have catalogued Spitzer protostellar
sources: a further 12 flows are tentatively associated with Spitzer
sources (IRS source numbers marked with a question mark in
Tables~\ref{smz} and \ref{dfs}).  Two well-defined flows, HH~1/2 and
V380~Ori-NE, have established sources that were not flagged as young
stars in the analysis of the Spitzer data (the former because it is
too faint; the latter because it is too close to a neighbouring
source); a third object, HH~83, was only observed in two of the four
Spitzer IRAC bands.  Closer examination of the mid-IR data shows that
HH~1/2 and V380~Ori-NE are probably driven by protostars, and HH~83 by
a disk excess source.

In total, 97 H$_2$ outflows (including all of the SMZ jets) were observed
with MAMBO at 1200\mic . 53 (55\%) are associated with MAMBO (MMS)
cores {\em and} young Spitzer sources; a further four outflow sources are
associated with extended 1200\mic\ emission or are unresolved from brighter,
nearby cores.

Of the 116 jets, 68 (59\%) exceed an arcminute in length (at a
distance of 450~pc, 1\arcmin\ is equivalent to 0.13~pc); 10 are
parsec-scale jets, exceeding 7.6\arcmin\ (SMZ sources 6, 49, 55, 67,
76 and the newly-discovered DFS flows 117, 123, 132, 133 and 138).
Throughout Orion~A we measure a range of jet lengths (as traced in
H$_2$); a histogram showing the distribution of lengths, uncorrected
for inclination angle, shows that the number of flows with a given
length decreases more-or-less exponentially (Fig.~\ref{histograms}a).
The same trend is seen if we just consider flows north of the ONC or
flows south of the ONC: the map in Fig.~\ref{lenmap} shows that the
distribution of short ($L < 1$\arcmin ), medium-length, and long ($L >
2$\arcmin ) jets is more-or-less even across Orion~A.  

The jet lengths in Tables~\ref{smz} and
\ref{dfs} are of course lower limits, since they are uncorrected for
inclination to the line of sight.  Also, as a jet or wind exits
its molecular cloud, it will no longer entrain and shock-excite
molecules and may not be visible in H$_2$ emission. The histograms in
Fig.~\ref{histograms}a are therefore biased towards shorter flow
lengths.  Even so, if all of the jets were of equal length and
randomly inclined with respect to the line of sight, inclination
alone would produce a histogram that increases towards longer jet
lengths (peaking at an angle of $\pi$/2, neglecting any selection effects
due to detectability as a function of angle).  We can therefore be
sure that the ``H$_2$ flows'' in Orion are mostly short; macro-jets are a
rare occurrence and, statistically at least, H$_2$ jet lengths do not
appear to be dictated by the changing environment in Orion~A.

We find no real correlation between H$_2$ jet length and indicators of the
outflow source age, such as the spectral index, $\alpha$ (defined in
Sect.~2.3), the integrated 1200\mic\ flux of the core associated with
each outflow source, $I_{1200}$, or the 1200\mic\ surface brightness
measured towards each outflow source, $F_{1200}$ (Fig.~\ref{params} -
note that $F_{1200}$ is not always the flux density measured towards
the core peak).  Five of the six longest H$_2$ flows are associated with IRS
sources with large spectral indices ($\alpha > 1.5$), though at the
same time four of these five protostars seem to be associated with
modest-sized cores ($I_{1200}< 1.5$~Jy; $F_{1200}< 0.6$~Jy) -- the
fifth protostar was not observed with MAMBO.  The longest H$_2$ jet in
Orion~A, SMZ~49 ($L
\sim$ 35\arcmin), is almost certainly driven by a source in the
compact L~1641-N cluster, which is associated with highly-reddened
protostars and dense, massive molecular cores. However, identifying
this remarkable jet's progenitor and, in particular, resolving the
associated core, is difficult.
 
There is some indication from millimetre observations of CO outflows
that Class 0 sources produce more collimated flows than their Class I
counterparts (\cite{lee02} 2002; \cite{arc06} 2006). This may be due
to the destruction of the protostellar core by the outflow itself, as
it entrains and plows away the ambient gas.  Since H$_2$ emission
traces the interaction of outflows with the surrounding molecular gas,
one might expect to see some indication of this ``flow broadening'' in
H$_2$ images of outflows from sources with lower values of $\alpha$,
or low values of $I_{\rm 1200}$. However, this does not seem to be the
case: the flow opening angle, defined as a cone that encloses all H$_2$
emission line features, with an apex centred on the outflow source,
appears to be unrelated to $\alpha$, $I_{\rm 1200}$ or $F_{\rm 1200}$
(Fig.~\ref{params}).

If we simply compare sources associated with a molecular core with
sources undetected in 1200\mic\ emission, we again find that both
samples essentially have the same distributions of H$_2$ jet lengths
and jet opening angles: for H$_2$ jet sources with cores the mean
length is 3.9\arcmin\ (standard deviation = 5.6\arcmin ; range =
0.2\arcmin --34\arcmin ); for those without cores the mean length is
1.7\arcmin\ (std dev = 1.9\arcmin ; range = 0.1\arcmin --9.5\arcmin ).
The former is skewed somewhat by the fact that four of the seven
longest H$_2$ flows ($L >$10\arcmin ) are associated with IRS sources that
do coincide with 1200\mic\ cores (the other three long flows were not
observed at 1200\mic ).  The mean opening angles for H$_2$ jets with and
without cores are 27.7\dg\ (std dev = 18\dg ; range = 4\dg --70\dg )
and 19.9\dg\ (std dev = 15\dg ; range = 3\dg --50\dg ), respectively.

The lack of a correlation between $\alpha$ and H$_2$ jet length, $L$,
or opening angle, $\theta$, may be a result of the fact that $\alpha$
varies with orientation as well as source ``age'' (e.g. \cite{whi03}
2003).  More evolved sources driving the longest jets are expected to
have lower spectral indices.  However, jets inclined towards the
observer will appear shorter and will likewise have lower spectral
indices.  Poor resolution and source confusion may also explain the
lack of a correlation between core parameters ($I_{\rm 1200}$ and
$F_{\rm 1200}$) and $L$ or $\theta$.  One should also note that H$_2$
is not always an ideal measure of flow parameters. As noted
earlier, H$_2$ excitation only occurs when the flow impacts dense
ambient material, and pre-existing H$_2$ features cool and fade very
rapidly (within a few years).  For a flow that exits a molecular
cloud, H$_2$ emission features will not be observed, so the true jet
length may be underestimated.  Conversely, jet opening angles may be
over-estimated: the wings of jet-driven bow shocks are often much
wider than the underlying jet (SMZ~42 is a nice example of this in
Orion), and changes in flow direction due to precession will, over
time, increase the apparent opening angle.

Finally, the precise relationship between $\alpha$ and source age
has not yet been established.  Spectral typing of protostars is
difficult because of extinction and veiling of photospheric absorption
lines (\cite{gre02} 2002; \cite{dop03} 2003; \cite{bec07} 2007;
\cite{ant08} 2008), so age estimates based on the fitting of
theoretical mass tracks and isocrones, especially for the youngest
sources, can be problematic (see e.g. \cite{dop05} 2005).  Radio
continuum observations may be a better way to quantify the youth of an
outflow source.

%%%%%%%%%%%
%%%%%%%%%%

\subsection{Flow orientations}
 
H$_2$ jet position angles on the sky should be well defined,
provided the outflow source is correctly identified. H$_2$ surveys can
therefore be used to test the standard paradigm that clouds collapse
along field lines to form elongated, clumpy filaments from which
chains of protostars are born (\cite{mou76} 1976; \cite{lar08} 2008),
with the associated outflows aligned parallel with the local field,
and perpendicular to the chains of cores (\cite{ban06} 2006).  The
presence of magnetic fields in molecular clouds has been demonstrated,
and the orientations of field lines with respect to chains of oblate
cores have been measured in a number of star forming regions
(e.g. \cite{cru99} 1999; \cite{mat02} 2002; \cite{hou04} 2004;
\cite{val06} 2006; \cite{lar08} 2008).  
In a few regions there is also some indication 
that outflows are preferentially orientated orthogonal to individual
cores or chains of cores, and/or parallel with the local magnetic
field (\cite{str86} 1986; \cite{hey87} 1987; \cite{men04} 2004;
\cite{dav07} 2007; \cite{ana08} 2008).

The Orion~A GMC comprises a large number of cores extended along a
north-south ``integral-shaped'' filament that passes through the Orion
nebula and extends southward, broadening out between L~1641-N and
L~1641-S (\cite{joh99} 1999, 2006; \cite{nut07} 2007).  It is
therefore potentially an ideal laboratory for an investigation into
the relationship between cores, filaments and outflows.  Throughout
much of Orion A, one might expect to see flows predominantly
orientated east-west.

In Fig.~\ref{histograms}c we have plotted a histogram of the jet
position angles (in ten degree bins) for all outflows in our sample.
The plot suggests a uniform distribution of position angles.
Moreover, a Kolmogorov-Smirnov (KS) test shows that there is a $>$99\%
probability that such a distribution is drawn from a homogeneously
distributed sample. In other words, the position angles of the H$_2$ flows
are completely random. If we consider only the jets with identified
outflow sources a similar situation is found. In this case the
probability that the distribution is random is 99.2\%. The same also
holds if we separate the H$_2$ jets in the northern region (Dec $>$ -6\dg ),
where the dense gas is more tightly constrained to a north-south
filament, from those south of the ONC (Dec $<$ -6\dg ).  Even for the
compact ridge of pre- and proto-stellar cores north of the ONC
(OMC~2/3), the outflows are randomly orientated. Similar plots for
H$_2$ jets exceeding an arcminute in length (flows that arguably have more
reliable position angles), or for jets with embedded protostars
coincident with MAMBO cores, are equally randomly orientated.

So the H$_2$ jets in Orion A appear to be randomly orientated.  Notably,
\cite{mat01} (2001) and \cite{hou04} (2004) find that on large scales
the field along the integral-shaped filament (north and south of the
ONC) is non-uniform. The long axes of the elongated cores
along the filament also appear to be randomly 
orientated (\cite{chi97} 1997; \cite{joh99} 1999, 2006;
\cite{nut07} 2007; see also e.g. Fig.~\ref{scu1}).  It is therefore not too
surprisingly that flow position angles should also be non-uniform. 
But are the H$_2$ outflows orientated orthogonal to the cores
associated with each outflow source? At 450~pc, the MAMBO beam of
11\arcsec\ corresponds to $5 \times 10^{3}$~AU, considerably larger
than accretion disk dimensions or the size of the solar system.  Even
so, many of the cores in Tables~\ref{smz} and \ref{dfs} are elongated;
if these ``circumstellar envelopes'' are magnetically tied to the
accretion disks they harbour (due to modest ambipolar diffusion), then
one might expect them to be orientated perpendicular to the flow axes
(\cite{ban06} 2006; \cite{ana08} 2008).

In our data, 47 H$_2$ jet sources are associated with 1200\mic\ emission
above a surface density $F_{1200} > 75$~mJy.  Of these, 42 are
associated with cores (rather than ``emission'' in Tables~\ref{smz}
and \ref{dfs}).  From this sample we extract sources where the core
major-to-minor axis ratio exceeds 1.3, and exclude very large
(diffuse) cores where the Full Width Half Maximum of the Gaussian fit
is greater than 50\arcsec\ in either axis.  We arrive at a sample of
just 22 H$_2$ flows with cores.  Surprisingly, even from this finely-tuned
sample, the distribution of H$_2$ jet position angles (PAs) with
respect to core PAs is completely random.  One might expect that the
``longer'', more clearly-defined jets, would be orthogonal to their
associated cores.  However, in a plot of jet length against ``jet-PA
minus core-PA'' (not shown), even the longest jets seem to be randomly
orientated with respect to their progenitor cores.

It seems likely that, in some of the more clustered regions (OMC~2/3,
L1641-N, etc.), the MAMBO observations have insufficient resolution to
disentangle the molecular envelopes associated with multiple
sources. Focusing on a hand-picked sample of sources separate from the
main clusters of protostars, sources with well-defined jets and
bright, compact cores, is probably necessary.  Obviously much higher
resolution (sub)mm observations would also be beneficial; ALMA will be
useful in this respect.

%%%%%%%%%%%%%%%%%%%%%%%%%%%%%%%%%%%%%%%%%%%%%%%%%%%%%%%%%%%%%%%%%
%%%%%%%%%%%%%%%%%%%%%%%%%%%%%%%%%%%%%%%%%%%%%%%%%%%%%%%%%%%%%%%%%

\section{Discussion}

 \begin{figure}
   \centering
   \includegraphics[width=85mm]
           {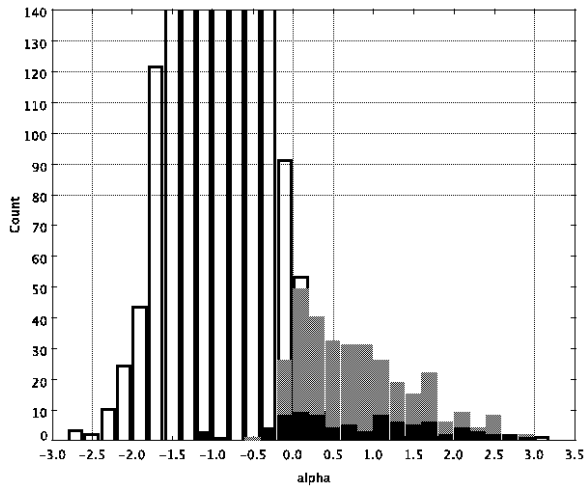}     
\caption {Histograms showing the number of disk excess sources (open
columns), protostars (filled grey columns) and H$_2$ flow sources
(black columns) with a given value of $\alpha$ (in bins of 0.2).}
    \label{alpha}
 \end{figure}

%%%%%%%%
%%%%%%%%

\subsection{Do all protostars drive H$_2$ outflows?}

In Fig.~\ref{alpha} we plot histograms showing the distribution of
spectral indices for the Spitzer disk excess sources, the Spitzer
protostars, and the outflow sources.  As was found to be the case in
Perseus (\cite{dav08} 2008) and the more distant, high mass star
forming region, W~75/DR~21 (\cite{dav07} 2007; \cite{kum07} 2007) the
majority of H$_2$ flows are driven by protostars with
positive spectral indices. The mean value of $\alpha$ for the disk
excess sources, most of which will be T Tauri stars, is -0.85
(std dev = 0.61); for the protostars and the outflow
sources, the mean values are 0.80 (std dev = 0.72) and
0.86 (std dev = 0.93); in terms of spectral index, the
outflow sources in Orion~A appear to be indistinguishable from the
protostellar sample.  

For the outflow sources in Perseus and DR~21/W~75 the mean values of
$\alpha$ are 1.4 and 1.9, respectively, somewhat higher than in Orion
A. The protostellar population in Orion~A may therefore be slightly
more evolved than in Perseus and DR~21/W~75. Note, however, that the
DR~21/W~75 spectral indices were derived from only IRAC photometry,
which may result in a higher mean value (the Perseus indices were
measured using 2MASS JHK photometry, as well as Spitzer and MIPS
data). Also, the larger UV radiation field from the ONC and from B-
and A-type stars spread throughout Orion A may accelerate the erosion
of protostellar envelopes, thereby reducing measured values of
$\alpha$, although $\alpha$ is largely defined by emission from the
inner disk rather than the outer, cold envelope, so this affect may be
modest. 

The spectral index is clearly a useful tool when searching for H$_2$
outflow source candidates.  Yet of the 290 Spitzer-identified
protostars observed in Orion~A with WFCAM (excluding the dozen sources
in the Orion Nebula), only $\sim$84, or less than one third, seem to
be driving outflows that excite H$_2$ emission.  Thus, at first
sight it does not appear that all protostars drive H$_2$ flows.

However, our near-IR images may not be sensitive enough to
detect the weakest flows, especially H$_2$ jets from the least luminous
-- and least massive -- protostars (\cite{car08} 2008). Also,
the selection criteria used for identifying the Spitzer
protostars may well include a handful of background galaxies and AGN
($\sim$0.7 per square degree), H$_2$ emission knots, and a number of more
evolved young stars.  Though faint, heavily reddened Class II sources
and disk excess sources viewed edge-on can have the same colours as
protostars (\cite{whi03} 2003).  These classical T Tauri stars rarely
produce extensive H$_2$ flows, because they have either swept away
much of their circumstellar environment, or because they occupy
regions with a low ambient molecular gas density (they may of
course drive HH jets, which our near-IR survey is not sensitive
to). \cite{gut09} (2009) estimate that at most 3.6\% of protostars
may be edge-on disk excess sources. Our WFCAM images include 1909 disk
sources (Megeath et al., in prep.).  There may therefore be as many of
70 disk sources in the protostar catalogue that are too evolved to
drive H$_2$ outflows.

The fact that the distributions of $\alpha$ values for the
outflow sources and protostars in Fig.~\ref{alpha} are comparable
suggests that protostars and H$_2$ outflow sources are essentially the
same objects; as noted earlier, the spread in $\alpha$ for both
samples may be due to inclination angle as much as to source age
(\cite{whi03} 2003; \cite{all04} 2004; \cite{meg04} 2004).  Moreover,
there is empirical evidence which suggests that the H$_2$ outflow
phase may be almost as long-lived as the protostellar phase: the
dynamical age of an H$_2$ flow is of the order of 10$^4$-10$^5$~years
(based on a canonical jet length of 1~pc and molecular outflow speeds
of 10-100~km s$^{-1}$; e.g. \cite{arc07} 2007), while statistical
studies indicate that the low mass protostellar phase may last at
least as long (\cite{fro06} 2006;
\cite{hat07} 2007; \cite{eno08} 2008).  

Towards the end of the protostellar phase the accretion rate -- and
therefore the mass loss rate -- will evolve quite dramatically, as the
source approaches the Class II T~Tauri phase.  Statistically at least,
these sources will probably be associated with mostly atomic, low
power jets, which may then persist for timescales of the order of
$10^5 - 10^6$~years (\cite{rei01} 2001). Such jets may be undetectable
without very deep H$_2$ imaging, though their sources will in any
case usually be identified as ``disk excess sources'' in the Spitzer
analysis used here.

In conclusion, it seems likely that for much of their evolution
{\em all protostars will produce outflows bright in H$_2$ line
emission}.  In a sample such as ours, the protostars not associated
with H$_2$ flows may be misclassified disk excess sources, AGN or even
H$_2$ knots. Some protostars may be marginally too evolved or too low
mass to produce detectable H$_2$ jets.  Orientation effects and the
local, inhomogeneous environment will also play a role in jet
detectability.

%%%%%%%%
%%%%%%%%

\subsection{Association of H$_2$ outflow sources with molecular cores}

In a recent study, \cite{jor07} (2007) used a combination of Spitzer
IRAC and MIPS photometry and 850\mic\ SCUBA observations to identify
the youngest protostars in Perseus.  They selected 49 MIPS sources that
have red [3.6]-[4.5] and/or [8.0]-[24] colours, and are within
15\arcsec\ of a SCUBA peak.  \cite{dav08} (2008) later found that most
(63\% of the reddened MIPS sources they imaged) drive H$_2$ flows.  In
Orion A it seems that a similarly high percentage of ``protostars with
cores'' drive molecular (H$_2$) outflows; note that most of the outflows
with identified progenitors in Tables~\ref{smz} and \ref{dfs}
are associated with discrete 1200\mic\ cores.

However, clearly not all molecular cores are associated with
H$_2$ flows.  We identify in excess of 500 peaks in the 1200\mic\
MAMBO data (Stanke et al., in prep.).  Similarly, \cite{nut07} (2007)
extract almost 400 cores from their more limited 850\mic\ SCUBA maps
of Orion~A.  Only about 1/5th of these cores seem to be
associated with H$_2$ outflows. \cite{joh06} (2006) obtain a similar
result in their comparison of SCUBA data with the H$_2$ images of
Sta02; they find that 17 of the 70 SCUBA clumps they identify are
clearly associated with H$_2$ emission.  

Many of the cores in Orion A will be prestellar -- based on the number
of cores without H$_2$ jets, roughly 80\% -- which suggests that the
H$_2$ outflow (and therefore protostellar) phase is considerably
shorter than the prestellar phase.  This statistic assumes that each
core produces just one protostar (and H$_2$ flow), which is almost
certainly incorrect; small groups of protostars are often found in
each core (see for example the L~1641-N core in Fig.~\ref{scu4}).
The prestellar phase could therefore be at least an order of
magnitude longer than the protostellar/H$_2$ outflow phase.

\begin{figure}
   \centering
   \includegraphics[width=85mm]
           {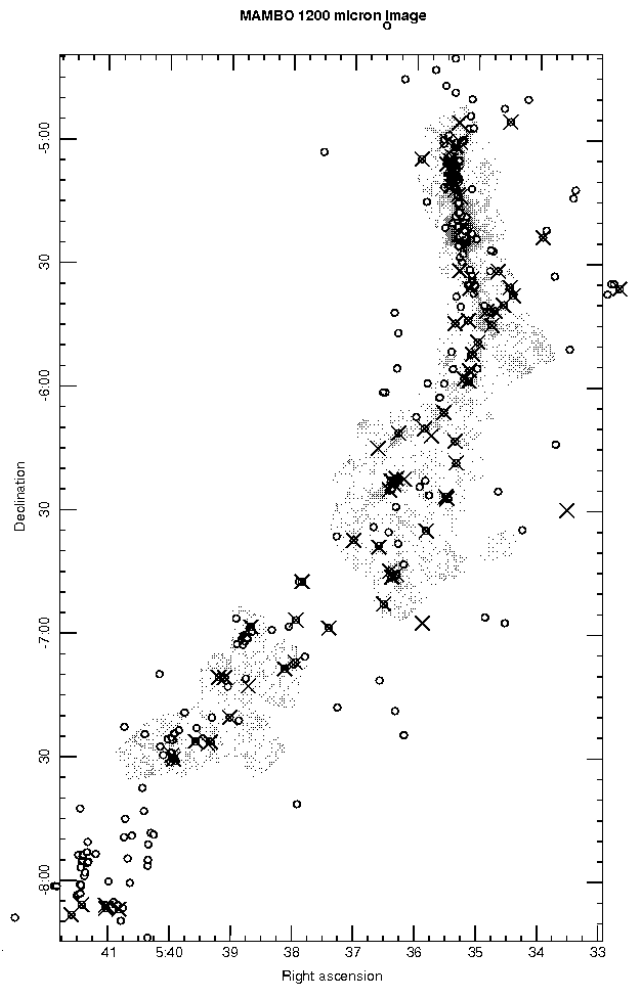} 
    \caption {A logarithmically-scaled image of 1200\mic\ emission
    with, superimposed, the locations of all Spitzer-identified
    protostars (open circles).  H$_2$ outflow sources are marked with
    crosses (note that a handful of H$_2$ jet sources are disk excess 
    sources).}
    \label{mambo}
 \end{figure}

What fraction of {\em cores containing protostars} drive H$_2$
flows?  In Fig.~\ref{mambo} we plot the entire 1200\mic\ map of
Orion~A.  We over-plot the positions of the Spitzer candidate
protostars, and mark the locations of the H$_2$ outflow sources.
(Note that the outflow and protostar catalogues are incomplete in the
Orion Nebula region; saturation effects, source confusion, and the
bright, diffuse nebulosity that pervades the ONC will limit both
samples.) Of the 190 protostars in Orion~A that were observed with
MAMBO, excluding the dozen in the Orion Nebula region, only 70
coincide with 1200\mic\ cores (to within an 11\arcsec\ radius),
although most do coincide with extended and/or diffuse emission.  Of
these 70 protostars with cores, 43 (61\%) appear to be driving
molecular H$_2$ outflows.  In some of the less tightly clustered and
therefore less complex regions in L~1641, such as the NGC~1980 region
in Fig.~\ref{scu3}, the number of protostars associated with cores
that also drive H$_2$ jets approaches 100\%.  Within the limitations
of our millimetre and near-IR observations (discussed in the previous
section), it seems likely that {\em most, if not all protostars within
molecular cores drive H$_2$ outflows.} It is also clear that the
combination of Spitzer mid-IR photometry and either SCUBA 850\mic\ or
MAMBO 1200\mic\ far-IR photometry is ideal for identifying molecular
outflow sources and, in all probability, the youngest protostars.

Lastly, we mention that, of the 30 or so protostars in
Fig.~\ref{mambo} that are located to the east or west of the edges of
the MAMBO map, i.e. beyond the bounds of the high-density molecular
ridge that constitutes OMC~2/3 and L~1641, only three protostars are
driving H$_2$ flows.  A much larger fraction of protostars
within the bounds of the MAMBO map, where ambient gas densities are
much higher, are driving molecular outflows.  As was noted in the
previous section, some of the protostars without H$_2$ jets may be
marginally more evolved; Fig.~\ref{mambo} suggests that these
``evolved protostars'', sources close to the Class II T Tauri phase
(which may of course drive HH jets), are more widely
distributed.

%%%%%%%%
%%%%%%%%

 \begin{figure}
%   \centering
   \hspace*{-0.5cm}
   \includegraphics[width=95mm]
           {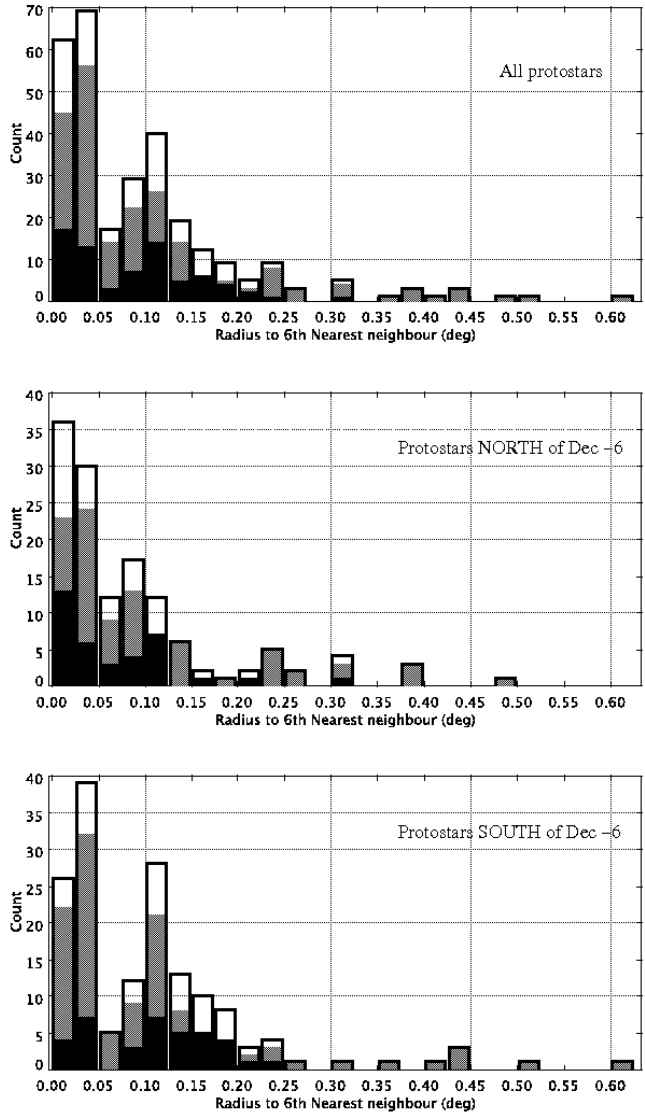} 
    \caption {Histograms showing the distribution of 6th nearest
    neighbour radii for all Spitzer-identified protostars (open bars),
    for those that drive H$_2$ outflows (black filled bars), and for those
    that do not drive H$_2$ outflows (grey filled bars). Top - entire region; 
    middle - north of declination -6\dg ; bottom - south of declination 
    -6\dg .}
   \label{nn}
 \end{figure}

%%%%%%%%%%%%%%%%%%%%%%%%%%
%% NN radii
%%%%%%%%%%%%%%%%%%%%%%%%%%
 
\begin{table*}[ht]
\centering
%\vspace*{-8.8cm}
\caption{Mean and median Nearest Neighbour (NN) radii for protostars
(all sources, sources with H$_2$ outflows, and sources without) in
Orion A. The standard deviation (std) is given in brackets. North and
South refer to protostars located north or south of declination -6\dg
, respectively.}
\label{mnn}
%\vspace*{-0.3cm}
\begin{tabular}{@{}l ccccc@{}}
   \hline\hline
   \noalign{\smallskip}
Region & 5th NN  & 6th NN & 7th NN & 10th NN & 20th NN \\
       & mean(std)median & mean(std)median & mean(std)median & mean(std)median & mean(std)median  \\
   \noalign{\smallskip}
   \hline
   \noalign{\smallskip}

All Protostars 		     & 0.087\dg(0.09\dg)0.060\dg & 0.097\dg(0.10\dg)0.070\dg & 0.106\dg(0.10\dg)0.079\dg & 0.13\dg(0.11\dg)0.11\dg & 0.20\dg(0.11\dg)0.18\dg \\
Protostars with flow	     & 0.077\dg(0.06\dg)0.064\dg & 0.089\dg(0.07\dg)0.085\dg & 0.098\dg(0.07\dg)0.096\dg & 0.13\dg(0.08\dg)0.12\dg & 0.19\dg(0.10\dg)0.20\dg \\  
Protostars no flow	     & 0.090\dg(0.10\dg)0.055\dg & 0.100\dg(0.11\dg)0.061\dg & 0.109\dg(0.11\dg)0.073\dg & 0.13\dg(0.12\dg)0.09\dg & 0.20\dg(0.12\dg)0.17\dg \\ \\

All Protostars (North)	     & 0.077\dg(0.08\dg)0.042\dg & 0.086\dg(0.09\dg)0.050\dg & 0.095\dg(0.10\dg)0.058\dg & 0.12\dg(0.11\dg)0.08\dg & 0.16\dg(0.12\dg)0.14\dg \\
Protostars with flow (North) & 0.059\dg(0.06\dg)0.035\dg & 0.068\dg(0.06\dg)0.048\dg & 0.075\dg(0.07\dg)0.058\dg & 0.10\dg(0.08\dg)0.11\dg & 0.14\dg(0.10\dg)0.14\dg \\  
Protostars no flow (North)   & 0.083\dg(0.09\dg)0.047\dg & 0.093\dg(0.10\dg)0.051\dg & 0.103\dg(0.11\dg)0.059\dg & 0.12\dg(0.12\dg)0.08\dg & 0.17\dg(0.13\dg)0.13\dg \\ \\

All Protostars (South)       & 0.096\dg(0.10\dg)0.075\dg & 0.107\dg(0.10\dg)0.094\dg & 0.115\dg(0.10\dg)0.101\dg & 0.14\dg(0.11\dg)0.12\dg & 0.23\dg(0.11\dg)0.21\dg \\
Protostars with flow (South) & 0.094\dg(0.06\dg)0.093\dg & 0.110\dg(0.06\dg)0.120\dg & 0.119\dg(0.06\dg)0.137\dg & 0.16\dg(0.07\dg)0.18\dg & 0.24\dg(0.06\dg)0.23\dg \\  
Protostars no flow (South)   & 0.097\dg(0.11\dg)0.073\dg & 0.106\dg(0.11\dg)0.079\dg & 0.114\dg(0.11\dg)0.083\dg & 0.14\dg(0.12\dg)0.11\dg & 0.23\dg(0.12\dg)0.19\dg \\

     \noalign{\smallskip}
      \hline
      \noalign{\smallskip}
   \end{tabular}
\end{table*}

%%%%%%%%
%%%%%%%%

\subsection{Clustering and the distribution of protostars and outflow
sources.}

Qualitatively speaking, if one examines Fig.~\ref{mambo} closely, it
appears that {\em clusters of protostars are not always associated
with clusters of H$_2$ outflows}. In L~1641-C, for example (RA $\sim 5^h
38.7^m$, Dec $\sim$ 7\dg 00\arcmin ; see also Fig.~\ref{scu8}),
although a dozen protostars and $\sim$30 disk excess sources are
identified, there is only one H$_2$ flow. Likewise, between Re~50 and
Haro~4-255 (RA $\sim 5^h 39.9^m$, Dec $\sim$ -7\dg 28\arcmin
; see also Fig.~\ref{scu10}) there is a chain of 16 protostars and at
least as many disk excess sources, yet again we only detect three or
four molecular H$_2$ outflows.  Neither region is associated with
large columns of ambient molecular material (L~1641-C is barely
detected in the MAMBO data); the paucity of detected outflows may
therefore be due to the evolved nature of the protostars in each
region.  L~1641-N, on the other hand, is associated with massive dense
cores, and a large fraction (roughly 50\%) of the protostars power 
H$_2$ outflows. Likewise, north of the ONC, in the OMC~2/3
molecular cloud region (between declinations -4\dg 30\arcmin\ and
-5\dg 15\arcmin) where protostars and cores are very highly clustered,
there are 65 protostars of which 26 appear to be driving molecular
flows (this may be a lower limit, given the abundance of H$_2$
emission line features, although protostars are also difficult to
resolve in this busy, nebulous region).

We also note that regions with multiple H$_2$ flows often seem
to be more sparsely populated with protostars.  For example, around
HH~1/2 (RA $\sim 5^h 36.3^m$, Dec $\sim$ -6\dg 46\arcmin; see also
Fig.~\ref{scu6}) there are at least half-a-dozen H$_2$ jets, yet only
about the same number of protostars.  Similarly, east of HH~1/2 around
DFS~123 (Fig.~\ref{scu7}) and south of HH~1/2 around HH~38/43/64
(Fig.~\ref{scu8}) the ratio of H$_2$ jets to protostars is
roughly one to one, though the number of protostars is again low.

However, neither of the above semi-quantitative observations seems to
be related to clustering. We demonstrate this by measuring the radius
to the 5th, 6th, 7th, 10th and 20th nearest neighbour (NN) for each of
the protostars identified in Orion A.  The distribution of 6th NN
radii is displayed in Fig.~\ref{nn}, where we distinguish protostars
that do drive H$_2$ flows from those that do not.  We also
separate protostars north of Dec -6\dg\ (the OMC~2/3 region) from
those south of Dec -6\dg\ (L~1641).  In Table~\ref{mnn} we list the
mean and median radius for each sample of Nth nearest neighbour radii,
and again distinguish outflow sources from the remaining sample of
protostars, and contrast the northern region with the south (median
values tend to exclude outliers with large radii located on the
periphery of each region).  Table~\ref{mnn} and Fig.~\ref{nn} both
demonstrate that there is no significant difference between the
spatial distribution of protostars that drive H$_2$ flows and
protostars that do not.  In other words, the molecular outflow
sources are no more -- nor no less -- clustered than the sources that
don't drive H$_2$ flows. This result is confirmed by K-S testing: for the
6th NN radii, in the northern region there is a 71\% probability that
the protostar and H$_2$-outflow-driving source radii are drawn from the same
sample; in the south the probability is lower (11\%) though still
inconclusive -- both distributions of NN radii are essentially the
same. K-S tests for the 5th and 7th NN radii yield similar results.

We conclude that the differences we note at the beginning of this
section are simply due to evolutionary effects.  There are clearly
regions within Orion A that are more evolved than others; young
regions like L~1641-N that contain embedded protostars that drive
H$_2$ outflows, and older regions like L~1641-C with clusters of
``protostars'' that are not associated with large clouds, dense cores or
molecular outflows.  The high fraction of H$_2$-outflow-driving
protostars suggests that transition time scales -- from protostar
with H$_2$ flow, via protostar without H$_2$ flow, to disk excess
source -- are relatively brief.  The fact that the protostars without
molecular outflows are no more widely distributed (on average) than
protostars with outflows suggests that the protostellar phase as a
whole is short in comparison to the overall time-scale over which
young stars become more widely distributed.

Finally, the NN analysis summarised in Table~\ref{mnn} also
demonstrates that there are considerable differences between OMC~2/3
and L~1641.  All protostars, regardless of whether they drive H$_2$ outflows
or not, are far more clustered in the north (NN radii are consistently
shorter) than in L~1641 in the south.  Clustering, the distributions
of protostars and disk excess sources, and the population of young
stars throughout Orion, will be discussed in detail in a forthcoming paper
(Megeath et al., in prep.).

%%%%%%%%%%%%%%%%%%%%%%%%%%%%%%%%%%%%%%%%%%%%%%%%
%%%%%%%%%%%%%%%%%%%%%%%%%%%%%%%%%%%%%%%%%%%%%%%%

\section{Summary and Conclusions}

We present near-IR images of H$_2$ outflows spread across an eight
square degree region that encompasses most of the Orion~A GMC.  For
much of this region Spitzer photometry is used to identify protostars
and disk excess sources, and large-scale (sub)millimetre maps are used
to plot the distribution of molecular gas.  Tangential velocities
(proper motions) are measured for features in 33 outflows; these data
are used to help associate the outflows with the protostars and dense
molecular cores.

We identify 43 new H$_2$ outflows, increasing the number of H$_2$
flows in Orion~A to 116 (not including the Orion bullets). From the
Spitzer sample of young stars we find sources for most of these
outflows.  Of the 300 or so protostars identified by Megeath et
al. (in prep.), at least 1/3 seem to be driving H$_2$ flows.  Indeed,
the H$_2$ outflow sources are predominantly protostars (Class 0/I
sources) rather than disk excess sources (mostly Class II T Tauri
stars), though a few of the latter source sample do seem to power
H$_2$ flows.  Most H$_2$ outflow sources have positive spectral
indices and are associated with dust cores; most molecular cores that
harbour protostars also drive molecular outflows.  All evidence points
to the extreme youth of H$_2$ jet progenitors.

We find no evidence for a preferred molecular outflow direction.  Even
in OMC~2/3, the region north of the ONC, where the molecular gas is
confined to a north-south chain of dense cores, the H$_2$ flows seem to
be randomly orientated.  Nor do we find a correlation between H$_2$ flow
length or opening angle and source spectral index or core flux at
1200\mic .  We find no evidence that ``more evolved'' sources,
i.e. those with low $\alpha$ or weak 1200\mic\ emission, drive longer
or less-well-collimated H$_2$ flows.  The caveats associated with
measuring flow parameters from H$_2$ observations are discussed in
light of these results.

Finally, we investigate the spatial distribution of H$_2$ outflow
sources in relation to the overall distribution of protostars.
Although well-known star clusters, such as L~1641-N and L~1641-C, are
associated with clusters of a dozen or so protostars and disk excess
sources, only groups or clusters of protostars that are still
surrounded by dense molecular gas seem to be associated with multiple
H$_2$ flows.  Protostars that power H$_2$ outflows are no more (nor no
less) clustered than protostars that do not.  This, together with the
fact that a high percentage of protostars do drive H$_2$ flows,
suggests that the H$_2$-outflow-driving phase of a protostar's early
evolution is only marginally shorter than the overall lifetime of the
protostar; it seems clear that H$_2$ jets fade very quickly as each
source evolves towards the Class II pre-main-sequence phase.

%%%%%%
%%%%%%

\section*{Acknowledgements}

We gratefully acknowledge the two referees, Bo Reipurth and John
Bally, for their careful reading of the text and very useful comments,
which served to improve the accuracy, clarity and overall quality of
the paper.  Thanks are also due to Dave Nutter for making the SCUBA
data available to us. We have made extensive use of the Starlink
analysis tools GAIA and TOPCAT (http://starlink.jach.hawaii.edu/).  We
acknowledge the Cambridge Astronomical Survey Unit (CASU) for
processing the near-IR data, and the WFCAM Science Archive in
Edinburgh for making the bulk of the WFCAM data discussed here
available to us.  The United Kingdom Infrared Telescope is operated by
the Joint Astronomy Centre on behalf of the U.K. Particle Physics and
Astronomy Research Council.  Some of the WFCAM data reported here were
obtained as part of the UKIRT Service Programme.  This research made
use of data products from the Spitzer Space Telescope Archive. These
data products are provided by the services of the Infrared Science
Archive operated by the Infrared Processing and Analysis
Centre/California Institute of Technology, funded by the National
Aeronautics and Space Administration and the National Science
Foundation.  We have also made much use of the SIMBAD database,
operated at CDS, Strasbourg, France.

%%%%%%%%%%%%%%%%%%%%%%%%%%%%%%%%%%%%%%%%%%%%%%%%%%%%%%%%%%%%%%
%%%%%%%%%%%%%%%%%%%%%%%%%%%%%%%%%%%%%%%%%%%%%%%%%%%%%%%%%%%%%%

%%%%%%%%%%%%%%%%%%%%%%%%%%%%%%%%%%%%%%%%%%%%%%%%%%%%%%%%%%%%%%%%%
%%%%%%%%%%%%%%%%%%%%%%%%%%%%%%%%%%%%%%%%%%%%%%%%%%%%%%%%%%%%%%%%%%
%%%%%%%%%%%%%%%%%%%%%%%%%%%%%%%%%%%%%%%%%%%%%%%%%%%%%%%%%%%%%%%%
%%%%%%%%%%%%%%%%%%%%%%%%%%%%%%%%%%%%%%%%%%%%%%%%%%%%%%%%%%%%%%%%%

%\newpage
\clearpage

%%%%%%%%%%%%%%%%%%%%%%%%%%
%% Old jets table
%%%%%%%%%%%%%%%%%%%%%%%%%%

\begin{sidewaystable}[h]
\centering
\vspace*{-8.8cm}
%\vspace*{9.2cm}
\caption{H$_2$ jets and outflows in Orion~A, from the catalogue of Sta02.}

\label{smz}
\vspace*{-0.3cm}

\begin{tabular}{@{}ccccc ccccc ccccc cc@{}}
   \hline\hline
   \noalign{\smallskip}
SMZ  & RA$^{\mathrm{a}}$    & Dec$^{\mathrm{a}}$           & Outflow$^{\mathrm{b}}$    & $\alpha^{\mathrm{b}}$ 
 & 850\mic $^{\mathrm{c}}$  & $Ap_{850}$$^{\mathrm{c}}$    & $I_{850}$$^{\mathrm{c}}$  &  $F_{850}$$^{\mathrm{c}}$
 & 1200\mic $^{\mathrm{d}}$ & $Ap_{1200}$$^{\mathrm{d}}$   & $I_{1200}$$^{\mathrm{d}}$ &  $F_{1200}$$^{\mathrm{d}}$
 & $L$$^{\mathrm{e}}$       & $\theta$$^{\mathrm{e}}$      & PA$^{\mathrm{e}}$         & HH$^{\mathrm{f}}$ 
  \\
& (J2000)  & (J2000) & source  & & peak & (arcsec) & (Jy) & (Jy/bm) & peak & (\arcsec ) & (Jy) & (Jy/bm) & (\arcmin ) & (deg) & (deg) &  \\
   \noalign{\smallskip}
   \hline
   \noalign{\smallskip}

1     & 5:35:19.3 & -4:55:45 & IRS~1  & -0.27 & no obs          &   --   & --  &  --   &undetected & --         &  --  &  --   & 0.3     & 20 & 68      &    -- \\ %Small bipolar jet\\
2     & 5:35:31.6 & -5:00:14 & IRS~2  & -1.04 & undetected      &   --   & --  &  --   &undetected & --         &  --  &  --   & 0.4     & -- & 140     &    -- \\ %Faint emission\\
3     & 5:35:18.3 & -5:00:33 & IRS~3  &  0.31 & AN-535182-50021 & 19x14  & 5.9 &  2.13 & MMS~3     & 22x14(92)  & 0.48 &  0.57 & 5.2     & 30 & 85      &    -- \\ %Knotty bipolar flow\\
4     & 5:35:23.5 & -5:01:29 & IRS~4  &  2.35 & AN-535235-50132 & 21x16  & 9.7 &  6.23 & MMS~4     & 137x50(147)& 4.64 &  1.93 & 0.5     & -- & 170     &    -- \\ %Compact H$_2$ knots \\
5     & 5:35:22.4 & -5:01:14 & IRS~5  &  2.90 & AN-535224-50114 & 13x10  & 3.9 &  2.60 & MMS~5     & 14x13(150) & 2.07 &  0.52 & 0.6     & 50 & 90      & 293   \\ %Knots $<$1\arcmin\ W of source\\
6     & 5:35:26.6 & -5:03:55 & IRS~6  &  1.53 & AN-535265-50356 & 37x28  & 6.5 &  1.63 & MMS~6     & 27x23(163) & 2.22 &  0.31 & 13      & -- & 85      &294/295\\ %Haro 5a/6a nebula \\
7     & 5:35:28.0 & -5:04:58 & IRS~7  &  1.04 & AN-535274-50511 & 32x23  & 5.3 &  0.44 & MMS~7     & 35x16(136) & 0.44 &  0.12 &$ \sim$6 & 10 & 30      &    -- \\ %Faint collimated jet \\
8     & (5:35:42.1& -5:04:39)&  ?     &  --   & ?               &   --   & --  &  --   & ?         & --         & --   &  --   &  --     & -- & --      &    -- \\ %Bow shock  \\
9     & 5:35:25.8 & -5:05:44 & IRS~9  &  1.54 & AN-535258-50551 & 38x22  & 6.3 &  1.55 & MMS~9     & 35x20(50)  & 1.31 &  0.24 & 2.8     & 42 & 80      & 330   \\ %Bright flow with knots \\
10    & 5:35:31.5 & -5:05:47 & IRS~10 & -0.12 & AN-535324-50547 & 50x28  & 5.3 &  0.42 & MMS~10    & 44x31(27)  & 2.22 &  0.12 & 1.5     & 15 & 50      & 287   \\ %Bow in bipolar flow  \\ 

11/13 & 5:35:23.3 & -5:07:10 & IRS~11 &  0.17 & AN-535236-50711 & 16x13  & 2.5 &  1.11 & MMS~11    & 31x18(122) & 1.08 &  0.20 & $\sim$1 & -- &$\sim$20 &    -- \\ %Curving chain of knots \\
12    & 5:35:27.7 & -5:07:04 & IRS~12 &  1.85 & AN-535279-50711 & 18x14  & 0.8 &  0.24 & MMS~12    & 29x23(173) & 0.54 &  0.11 & 0.3     & -- & 8       &    -- \\ %Small bipolar jet  \\
14/16 & 5:35:25.6 & -5:07:57 & IRS~14 &  0.93 & emission        &   --   & --  &  0.43 & MMS~14    & 20x15(47)  & 0.27 &  0.06 & $\sim$4 & 35 & 40      & 383   \\ %Bipolar outflow \\
15    & 5:35:22.4 & -5:08:05 & IRS~15 & -0.10 & AN-535222-50817 & 18x14  & 1.2 &  0.38 & MMS~15    & 20x15(98)  & 0.29 &  0.09 & 0.4     & -- &$\sim$45 & 385   \\ %Conical nebula $+$ knot \\
17    & 5:35:27.0 & -5:09:54 & IRS~17?&  0.54 & emission        &   --   & --  &  5.80 & MMS~17    & 25x22(41)  & 6.06 &  1.07 & 1.8     & 50 & 25      & 887   \\ %Bright cavity and bow \\
18    &(5:35:29.6 & -5:08:58)&   ?    &  --   & ?               &   --   & --  &  --   & ?         & --         & --   &  --   &  --     & 50 & --      &    -- \\ %Part of SMZ~17? \\
19    & 5:35:26.8 & -5:09:25 & IRS~19 & -0.03 & emission        &   --   & --  &  1.90 & MMS~19    & 20x15(92)  & 0.87 &  0.28 & 0.5     & -- & 100     & 384   \\ %Ribbed conical nebula \\
20    & 5:35:25.7 & -5:09:49 & IRS~20 &  0.14 & emission        &   --   & --  &  1.45 & emission  & --         & --   &  0.36 & 0.7     & -- & 175     &    -- \\ %Knots in complex region \\ 

21/22 & 5:35:26.9 & -5:11:07 & IRS~21 & -1.01 & emission        &   --   & --  &  0.33 & emission  & --         & --   &  0.10 & 1.6     & -- & 175     &    -- \\ %Two elongated knots \\
23    & 5:35:24.7 & -5:10:30 & IRS~23 &  1.18 & emission        &   --   & --  &  0.53 & MMS~23    & 20x16(175) & 0.48 &  0.18 &$\sim$4  & -- & 17      &    -- \\ %Bow/chain of knots \\
24    & 5:35:23.3 & -5:12:03 & IRS~24 &  1.37 & AN-535234-51205 & 21x16  & 3.2 &  1.28 & MMS~24    & 16x12(174) & 0.44 &  0.35 & 1.5     & 27 & 61      & 536   \\ %Conical nebular $+$ bows\\
25    & 5:35:20.1 & -5:13:16 & IRS~25 &  0.04 & emission        &   --   & --  &  0.87 & MMS~25    & 23x12(126) & 0.65 &  0.29 & 3.0     & -- & 160     & 44/535\\ %Elongated knot $+$ bows\\ 
26    &(5:35:10.9 & -5:23:12)&   ?    & --    & ?               &   --   & --  &  --   & ?         & --         & --   &  --   & 2.6     & -- & 125     &    -- \\ %Features W of BN/KL \\
27    &(5:35:11.6 & -5:23:41)&   ?    & --    & ?               &   --   & --  &  --   & ?         & --         & --   &  --   &  --     & -- & --      & 510?  \\ %Features W of BN/KL \\
28    &(5:35:10.9 & -5:23:46)&   ?    & --    & ?               &   --   & --  &  --   & ?         & --         & --   &  --   &  --     & -- & --      & 510?  \\ %Features W of BN/KL \\
29    & 5:34:40.9 & -5:31:44 & IRS~29 &  1.47 & no obs          &   --   & --  &  --   & MMS~29    & 21x16(160) & 0.42 &  0.16 & 5.0     & 20 & 156     &    -- \\ %Conical nebula $+$ knots\\
30    & 5:35:18.3 & -5:31:42 & IRS~30 & -0.36 & undetected      &   --   & --  &  --   & MMS~30    & 18x17(64)  & 0.06 &  0.03 & 0.9     & -- & 25      & 540   \\ %Faint knots \\ 

31    & 5:34:35.5 & -5:39:59 & IRS~31 &  1.01 & no obs          &   --   & --  &  --   & MMS~31    & 33x25(22)  & 0.26 &  0.06 & 0.9     & 60 & 150     &    -- \\ %Small bipolar flow \\
32    &(5:35:11.4 & -5:39:38)& ?      & --    & ?               &   --   & --  &  --   & ?         & --         & --   &  --   & 0.8     & -- & 170     &    -- \\ %Faint knots \\
33    & 5:34:51.9 & -5:41:33 & IRS~33?& 1.64  & undetected      &   --   & --  &  --   &undetected & --         & --   &  --   & 0.5     & -- & 134     &    -- \\ %Two knots $\sim$30\arcsec\ apart \\
34    & 5:35:09.9 & -5:43:45 & IRS~34 & 1.54  & undetected      &   --   & --  &  --   &undetected & --         & --   &  --   & --      & -- & --      &    -- \\ %Compact knots \\
35    & 5:34:46.9 & -5:44:51 & IRS~35?& 0.73  & undetected      &   --   & --  &  --   &undetected & --         & --   &  --   & 1.0     & -- & 58      &    -- \\ %Faint knot \\
36    &(5:35:10.0 & -5:45:06)& ?      & --    & ?               &   --   & --  &  --   & ?         & --         & --   &  --   & --      & -- & --      &    -- \\ %Small knots \\  
37    & 5:35:05.5 & -5:51:54 & IRS~37 & 0.77  & no obs          &   --   & --  &  --   &undetected & --         & --   &  --   & --      & -- & --      &    -- \\ %Small group of knots \\
38    & 5:35:08.6 & -5:55:54 & IRS~38 & 1.80  & no obs          &   --   & --  &  --   & MMS~38    & 19x15(121) & 0.88 &  0.31 &$\sim$7  & 10 &$\sim$150&    -- \\ %Curving, bipolar flow\\
39    & 5:35:09.0 & -5:58:28 & IRS~39 & 1.06  & emission        &   --   & --  &  0.41 & MMS~39    & 19x15(155) & 0.54 &  0.17 &$>$3     & 20 & $\sim$30&    -- \\ %Knots NE of source \\
40    & 5:35:33.2 & -6:06:10 & IRS~40?& 0.46  & emission        &   --   & --  &  --   &undetected & --         & --   &  --   & 3.8     & -- & 16      &    -- \\ %Bow 4\arcmin\ NNE of source\\ 

41    & 5:35:13.4 & -5:57:58 & IRS~41 & 2.30  & no obs          &   --   & --  &  --   & MMS~41    & 19x14(153) & 0.79 &  0.21 & 6.0     & -- & 9       &    -- \\ %Bright, knotty bow \\
42    & 5:35:22.2 & -6:13:06 & IRS~42 & 1.21  & AS-535213-61313 & 26x17  & 0.7 &  0.14 & MMS~42    & 57x37(35)  & 0.88 &  0.06 & 5.0     & 12 & 69      &    -- \\ %Collimated jet $+$ bows \\
43    & 5:35:52.0 & -6:10:02 & IRS~43 & 0.37  & AS-535524-61007 & 33x19  & 1.3 &  0.49 & MMS~43    & 22x16(150) & 0.58 &  0.19 &$\sim$2  & -- &$\sim$45 &    -- \\ %PDR around bright star?\\
44    & 5:35:45.2 & -6:11:44 & IRS~44?& -1.19 & undetected      &   --   & --  &  --   &undetected & --         & --   &  --   & 1.3     & -- & 132     &    -- \\ %Knot with cavity/bow  \\
45    & 5:36:17.3 & -6:11:11 & IRS~45 & 0.93  & undetected      &   --   & --  &  --   & MMS~45    & 32x22(166) & 0.22 &  0.04 & 2.3     & 60 & 3       &    -- \\ %Curving bipolar flow \\

     % \noalign{\smallskip}
      \hline
      \noalign{\smallskip}
   \end{tabular}
\end{sidewaystable}

\clearpage
\begin{sidewaystable}
\centering
%\vspace*{-8.5cm}
\vspace*{9.6cm}
\begin{tabular}{@{}ccccc ccccc ccccc cc@{}}
   \hline
   \noalign{\smallskip}
46    & 5:36:37.0 & -6:14:58 & IRS~46 & 0.51  & AS-536366-61459 & 19x13  & 0.8 &  0.30 & MMS~46    & 29x20(128) & 0.52 &  0.11 &$\sim$0.5& -- &$\sim$0  & 304   \\ %Bipolar nebula$+$flow\\
47    &(5:35:38.1 & -6:15:08)& ?      & --    & ?               &   --   & --  &  --   & ?         & --         & --   &  --   & 0.9     & -- & 155     &    -- \\ %Knots and bow shock \\ 
48    & 5:36:10.5 & -6:19:55 & ?      & --    & ?               &   --   & --  &  --   & ?         & --         & --   &  --   & 1.4     & -- & 40      & 299   \\ %Filament NE of source\\
49    & 5:36:19.5 & -6:22:12 & IRS~49 & 0.79  & AS-536196-62209 & 77x44  & 15.8&  1.92 & MMS~49    & 20x9(88)   & 0.16 &  0.53 &$\sim$34 & -- & 175     & 303   \\ %Big flow from L1641-N \\
50    & 5:36:11.5 & -6:22:22 & IRS~50?& -0.90 & undetected      &   --   & --  &  --   &undetected & --         & --   &  --   & 2.0     & -- &$\sim$170&    -- \\ %Chain of knots \\ 

51    & 5:36:24.6 & -6:22:41 & IRS~51 & 1.23  & AS-536250-62241 & 49x27  & 3.1 &  0.59 & MMS~51    & 52x24(155) & 1.39 &  0.12 &$>$4.0   & 40 & 69      & 301   \\ %Bows E of L1641-N  \\
52    &(5:36:39.1 & -6:22:38)& ?      & --    & ?               &   --   & --  &  --   & ?         & --         & --   &  --   &  --     & -- & --      &    -- \\ %faint bow in DFS115? \\
53    & 5:36:18.8 & -6:22:10 & IRS~53 & 2.47  & AS-536196-62209 & 77x44  & 15.8&  3.72 & MMS~53    & 17x16(38)  & 3.09 &  1.01 & 0.6     & -- & 45      &    -- \\ %Knots in L1641-N cluster\\
54    &(5:36:31.0 & -6:24:49)& ?      & --    & ?               &   --   & --  &  --   & ?         & --         & --   &  --   &  --     & -- & --      &    -- \\ %Faint knots \\
55    & 5:35:29.8 & -6:26:59 & IRS~55 & 1.77  & AS-535297-62701 & 27x19  & 1.7 &  0.83 & MMS~55    & 14x13(4)   & 1.92 &  0.26 & 17      & -- & 165     & 34    \\ %Major HH flow  \\
56    & 5:35:30.9 & -6:26:32 & IRS~56 & 1.16  & AS-535305-62622 & 32x21  & 1.6 &  0.26 & MMS~56    & 54x38(168) & 1.34 &  0.09 & 1.8     &  4 & 8       &    -- \\ %Collimated jet nr HH~34 \\
57    &(5:36:56.9 & -6:34:17)& ?      & --    & ?               &   --   & --  &  --   & ?         & --         & --   &  --   &  --     & -- & --      & 292   \\ %Knots by star \\
58    & 5:37:00.5 & -6:37:11 & IRS~58 & -0.16 & AS-537004-63715 & 36x19  & 1.5 &  0.44 & MMS~58    & 63x38(179) & 2.22 &  0.15 & 4.0     & 12 & 60      &    -- \\ %Bipolar jet/large bows \\
59    & 5:36:36.1 & -6:38:52 & V380~Ori-NE&-- & AS-536361-63857 & 38x18  & 1.8 &  0.93 & MMS~59    & 38x18(167) & 1.62 &  0.35 & 2.0     & 18 &$\sim$0  &    -- \\ %Bipolar H$_2$/CO flow \\
60    &(5:36:22.0 & -6:41:42)& ?      & --    & ?               &   --   & --  &  --   & ?         & --         & --   &  --   &  --     & -- & --      & 35    \\ %Knots nr V380~Ori \\ 

61    & 5:36:18.9 & -6:45:23 & IRS~61?& 2.07  & emission        &  --    & --  &  0.88 & emission  & --         & --   &  0.25 &$\sim$3.6& -- &$\sim$165&    -- \\ %Curving flow \\
62    &(5:36:11.6 & -6:43:04)& ?      & --    & ?               &  --    & --  &  --   & ?         & --         & --   &  --   &  --     & -- & --      & 3     \\ %Extension of HH~1/2? \\
63    & 5:36:25.1 & -6:44:42 & IRS~63 & 0.04  & AS-536259-64440 & 38x27  & 2.2 &  0.47 & MMS~63    & 58x48(20)  & 2.91 &  0.17 & 0.5     & 36 &$\sim$45 & 147   \\ %Knots 20\arcsec\ SW of source \\
64    & 5:36:22.8 & -6:46:06 &HH~1/2-VLA& --  & AS-536229-64618 & 52x30  & 6.2 &  1.36 & MMS~64    & 42x27(25)  & 5.61 &  0.42 &$>$3     & 32 &$\sim$145& 1/2   \\ %Bipolar H$_2$/HH flow \\
65    &(5:36:21.2 & -6:46:08)& ?      & --    & ?               &  --    & --  &  --   & --        & --         & --   &  --   & 1.9     & -- & 110     & 144   \\ %Small flow near HH~1/2\\
66    & 5:38:40.5 & -6:58:22 & IRS~66 & -0.04 & no obs          &  --    & --  &  --   & MMS~66    & 13x13(168) & 0.93 &  0.19 & 2.6     &  4 & 130     &    -- \\ %Collimated, knotty jet\\
67    & 5:37:57.0 & -7:06:56 & IRS~67 & 2.07  & no obs          &  --    & --  &  --   & MMS~67    & 22x20(98)  & 0.79 &  0.14 & 16      & 18 & 126     &38/43/64\\ %Bright, bipolar flow \\
68    & 5:38:07.5 & -7:08:29 & IRS~68 & 0.25  & no obs          &  --    & --  &  --   & MMS~68    & 24x17(2)   & 0.12 &  0.02 & 1.2     & 10 & 130     &    -- \\ %Parallel to HH~38/43 \\
69    & 5:39:05.8 & -7:10:39 & IRS~69 & 2.63  & no obs          &  --    & --  &  --   &undetected & --         & --   &  --   &$\sim$0.4& -- & 90      &    -- \\ %Bow 20\arcsec\ W of source \\
70    & 5:38:42.8 & -7:12:44 & IRS~70 & -0.23 & no obs          &  --    & --  &  --   & MMS~70    & 15x11(145) & 0.07 &  0.04 &$\sim$1.7& 30 &$\sim$135& 449   \\ %Faint bipolar flow \\ 

71    & 5:39:00.9 & -7:20:23 & IRS~71 & 1.55  & no obs          &  --    & --  &  --   & MMS~71    & 65x31(173) & 0.40 &  0.01 & 0.9     & -- & 128     &    -- \\ %Faint jet \\
72    & 5:39:19.6 & -7:26:19 & IRS~72b& 0.33  & no obs          &  --    & --  &  --   & MMS~72    & 22x20(176) & 3.17 &  0.30 & 5.4     & 70 & 65      & 469   \\ %Haro~4-255~FIR jet \\
73    & 5:39:22.3 & -7:26:45 & IRS~73 & -0.37 & no obs          &  --    & --  &  --   & MMS~73    & 15x10(24)  & 0.34 &  0.09 & 2.3     & 14 & 133     & 470   \\ %Bow from T~Tauri star \\
74    &(5:40:23.7 & -7:20:32)& ?      & --    & no obs          &  --    & --  &  --   & ?         & --         & --   &  --   &$>$1.0   & -- & 53      &    -- \\ %Small Knots \\
75    &(5:40:25.8 & -7:22:14)& ?      & --    & no obs          &  --    & --  &  --   & ?         & --         & --   &  --   &$>$1.5   & -- & 45      &    -- \\ %Chain of bows  \\
76    & 5:39:55.9 & -7:30:28 & IRS~76 & 2.70  & no obs          &  --    & --  &  --   & MMS~76    & 16x15(117) & 2.62 &  0.55 & 14      & -- &$\sim$60 &  65   \\ %Large curving flow \\ 

     % \noalign{\smallskip}
      \hline
      \noalign{\smallskip}

   \end{tabular}
\vspace*{-0.2cm}
\begin{list}{}{}

\item[$^{\mathrm{a}}$]Position of the H$_2$ outflow source if one is
listed in column 4.  Otherwise, the position of the brightest knot in
the flow is given (enclosed in brackets). Note that the well-known
sources of the HH~1/2, HH~83 and V380-Ori-NE flows were not identified in
our tables of Spitzer protostars or disk excess sources.

\item[$^{\mathrm{b}}$]Most likely H$_2$ outflow source from the Spitzer
photometry together with the spectral index, $\alpha$.  Just a
question mark means there is no obvious H$_2$ flow source candidate.

\item[$^{\mathrm{c}}$]850\mic\ dust core coincident with the H$_2$ outflow
source (from \cite{nut07} 2007).  The source must lie within a
14\arcsec\ radius (the JCMT beam at 850\mic ) of the core position
given by Nutter \& Ward-Thompson in their Table A1.  ''Emission'' or
``undetected'' means that no core appears in their table.  However,
``emission'' means that the source identified in column 4 is
associated with diffuse 850\mic\ emission (surface brightness
$>$100~mJy~beam$^{-1}$, which is roughly equivalent to 5$\sigma$); a
question mark means there are cores (or emission) in the vicinity
which could be associated with the outflow or its source; ``no obs''
means that the outflow is outside the bounds of the SCUBA map.  The
core size (major/minor axis dimensions), $Ap_{850}$, and integrated
flux, $I_{850}$, are given for each core; $F_{850}$ is the flux, in
Jy~beam$^{-1}$, measured towards each IRS source (not necessarily the
peak flux of the associated core).

\item[$^{\mathrm{d}}$]Same as for the 850\mic\ cores, but for cores
identified from our analysis of the 1200\mic\ observations of Stanke
et al. (in prep.). In this case the source must lie within 11\arcsec\
of the core centroid; ``emission'' corresponds to a
surface brightness $>$75~mJy~beam$^{-1}$ ($\sim 5\sigma$). The values
given in brackets in the $Ap_{1200}$ column are the orientation of the core
major axis (this information was not available for the 850\mic\
cores).

\item[$^{\mathrm{e}}$]Entire length ($L$) of all H$_2$ knots (in both
lobes), or the distance from the source to the most distant H$_2$ knot if
only one flow lobe is identified; opening angle ($\theta$) measured
from a cone centred on the outflow source that includes all H$_2$
features in the flow; position angle (PA) measured east of north.

\item[$^{\mathrm{f}}$]Associated HH objects, if any are known.

\end{list}
\end{sidewaystable}

%%%%%%%%%%%%%%%%%%%%%%%%%%
%% New jets table
%%%%%%%%%%%%%%%%%%%%%%%%%%
 
\clearpage

\begin{sidewaystable}
\centering
%\vspace*{9.5cm}
\vspace*{-9.2cm}
\caption{New H$_2$ jets in Orion~A (additions to the catalogue of Sta02). 
Columns are the same as for Table~\ref{smz}. }

\label{dfs}
\vspace*{-0.3cm}

\begin{tabular}{@{}ccccc ccccc ccccc cc@{}}
   \hline\hline
   \noalign{\smallskip}
DFS  & RA$^{\mathrm{a}}$    & Dec$^{\mathrm{a}}$         & Outflow$^{\mathrm{b}}$    & $\alpha^{\mathrm{b}}$ 
 & 850\mic $^{\mathrm{c}}$  & $Ap_{850}$$^{\mathrm{c}}$  & $I_{850}$$^{\mathrm{c}}$  &  $F_{850}$$^{\mathrm{c}}$
 & 1200\mic $^{\mathrm{d}}$ & $Ap_{1200}$$^{\mathrm{d}}$ & $I_{1200}$$^{\mathrm{d}}$ &  $F_{1200}$$^{\mathrm{d}}$
 & $L$$^{\mathrm{e}}$       & $\theta$$^{\mathrm{e}}$    & PA$^{\mathrm{e}}$         & HH$^{\mathrm{f}}$ 
  \\
& (J2000)  & (J2000) & source  & & peak & (arcsec) & (Jy) & (Jy/bm) & peak & (\arcsec ) & (Jy) & (Jy/bm) & (\arcmin ) & (deg) & (deg)  \\
   \noalign{\smallskip}
   \hline
   \noalign{\smallskip}

101  & 5:34:29.5 & -4:55:31  & IRS~101 & 1.37   &  no obs         &  --   &  --  &  --   & no obs    &  --        & --   &  --   &  5.3    &  --  & $\sim$160&    -- \\ %Curving, bipolar jet \\
102  & 5:35:55.7 & -5:04:38  & IRS~102 & 0.04   & undetected      &  --   &  --  &  --   & no obs    &  --        & --   &  --   & $>$3.5  &   9  & $\sim$90 & 42    \\ %Bow shocks \\
103  & 5:35:24.3 & -5:08:31  & IRS~103 & 1.18	& AN-535245-50832 & 26x16 &  4.3 &  1.55 & MMS~103   & 21x16(178) & 1.06 &  0.35 &  0.2    &  40  & $\sim$170&    -- \\ %Small knot nr source  \\
104  & 5:35:18.5 & -5:13:38  & IRS~104 & -0.05  & AN-535183-51338 & 13x8  &  0.3 &  0.07 & MMS~104   & 18x12(180) & 0.12 &  0.10 &  0.25   &  --  & 135      &    -- \\ %Small bow nr source \\
105  & 5:33:57.4 & -5:23:30  & IRS~105 & 2.34   & no obs          &  --   &  --  &  --   & no obs    &  --        & --   &  --   &  1.7    &  30  & 11       &    -- \\ %Curving flow \\
106  & 5:34:29.4 & -5:35:43  & IRS~106 & 1.78   & no obs          &  --   &  --  &  --   & MMS~106   & 109x20(160)& 0.31 &  0.01 &$\sim$0.5&  --  & $\sim$10 &    -- \\ %Faint jet S of source \\
107  & 5:34:26.4 & -5:37:41  & IRS~107 & 0.06   & no obs          &  --   &  --  &  --   &undetected &  --        & --   &  --   &  1.5    &   3  & 95       &    -- \\ %Faint jet E of source \\
108  & 5:35:06.5 & -5:33:35  & IRS~108?& 0.43   & emission        &  --   &  --  &  0.16 & MMS~108   & 18x7(114)  & 0.07 &  0.08 &  0.5    &  --  & $\sim$95 &541/882\\ %Knots W of source \\
109  & 5:35:08.5 & -5:35:59  & IRS~109 & 1.21   & AN-535082-53558 & 34x25 & 3.3  &  0.88 & MMS~109   & 74x45(20)  & 4.83 &  0.19 &  4.7    &   5  & 93       &    -- \\ %Bows E of source \\ 
110  & 5:32:42.5 & -5:35:55  & IRS~110 & 0.87	& no obs          &  --   &  --  &  --   & no obs    &  --        & --   &  --   &  --     &  --  & --       &    -- \\ %Knots nr Haro~4-145\\ 

111  & 5:34:44.1 & -5:41:26  & IRS~111?& 2.10   & AS-534439-54128 & 21x20 & 0.8  &  0.28 & MMS~111   & 34x28(115) & 0.71 &  0.09 &  0.8    &  25  & $\sim$50 &    -- \\ %Bow NE of source \\
112  & 5:35:22.6 & -5:44:30  & IRS~112 & 0.26  	& undetected      &  --   &  --  &  --   &undetected &  --        & --   &  --   &  3.0    &   4  & 105      &    -- \\ %Bow E of source \\
113  & 5:35:00.5 & -5:49:02  & IRS~113?& 0.61   & no obs          &  --   &  --  &  --   &undetected &  --        & --   &  --   &  3.7    &  11  & 5        &    -- \\ %Faint bow S of source\\
114  &(5:32:22.0 & -5:34:52) &  ?      & --     & no obs          &  --   &  --  &  --   & no obs    &  --        & --   &  --   &  --     &  --  & --       &    -- \\ %Group of faint knots \\
115  & 5:36:25.9 & -6:24:58  & IRS~115 & 0.13   & emission        &  --   &  --  &  0.73 & emission  &  --        & --   &  0.20 & $>$1.2  &  10  & $\sim$60 &    -- \\ %Bows NE of L1641-N \\
116  & 5:35:20.9 & -6:18:22  & IRS~116?& 0.18	& undetected      &  --   &  --  &  --   &undetected &  --        & --   &  --   &  1.3    &  29  & 133      & 33/40 \\ % Well-known HH objects \\
117  & 5:33:32.5 & -6:29:44  &HH~83-IRS& --     & no obs          &  --   &  --  &  --   & no obs    &  --        & --   &  --   & 10.8    &   6  & 108      & 83/84 \\ %HH jet and distant bows\\
118  & 5:36:21.8 & -6:23:30  & IRS~118 & -0.08	& AS-536220-62325 & 19x17 & 0.6  &  0.29 & MMS~118   & 15x11(89)  & 0.17 &  0.11 &  0.35   &  35  & 165      &    -- \\ %Small bows S of source \\
119  & 5:35:50.0 & -6:34:53  & IRS~119 & 1.37   & emission        &  --   &  --  &  0.16 &undetected &  --        & --   &  --   &$\sim$0.1&  --  & $\sim$80 &    -- \\ %Small knot by source \\
120  &(5:37:47.0 & -6:53:33) &   ?     & --     & no obs          &  --   &  --  &  --   & no obs    &  --        & --   &  --   &$\sim$0.1&  --  & $\sim$120&    -- \\ %Two compact knots  \\ 

121  & 5:36:23.5 & -6:46:15  & IRS~121 & 0.79   & AS-536229-64618 & 52x30 & 6.2  &  0.96 & MMS~121   & 41x27(25)  & 5.61 &  0.25 &  6.0    &   4  & 71       & 36    \\ %Bows E of HH~1/2 \\
122  & 5:37:51.0 & -6:47:20  & IRS~122 & 1.07	& no obs          &  --   &  --  &  --   & no obs    &  --        & --   &  --   &  3.0    &  33  & 142      & 89    \\ %Bipolar flow \\
123  &(5:37:33.1 & -6:50:20) & ?       & --     & no obs          &  --   &  --  &  --   & no obs    &  --        & --   &  --   &  8.5    &  30  & $\sim$95 &    -- \\ %Large bipolar jet \\
124  & 5:36:31.0 & -6:52:41  & IRS~124 & 0.39	& no obs          &  --   &  --  &  --   &undetected &  --        & --   &  --   &  1.0    &  39  & $\sim$140&    -- \\ %Bow SE of source \\
125  &(5:35:53.2 & -6:57:13) & no obs  & --     & no obs          &  --   &  --  &  --   & no obs    &  --        & --   &  --   &  3.7    &  --  & $\sim$15 & 127   \\ %Collimated, knotty jet \\
126  & 5:37:24.5 & -6:58:33  & IRS~126 & -0.02	& no obs          &  --   &  --  &  --   & no obs    &  --        & --   &  --   &  5.6    &   6  & 120      &    -- \\ %Bow NW of source \\
127  & 5:37:56.6 & -6:56:39  & IRS~127 & 2.02 	& no obs          &  --   &  --  &  --   & no obs    &  --        & --   &  --   &$\sim$0.1&  40  & $\sim$90 &    -- \\ %Knot by source \\
128  &(5:37:59.0 & -7:13:15) &   ?     & --     & no obs          &  --   &  --  &  --   & ?         &  --        & --   &  --   &  --     &  --  & --       &    -- \\ %Two faint knots \\
129  & 5:39:11.9 & -7:10:35  & IRS~129 & 0.25   & no obs          &  --   &  --  &  --   & MMS~129   & 59x35(154) & 1.04 &  0.03 &$\sim$0.8&  --  & $\sim$140&    -- \\ %Knot SE of source \\
130  &(5:40:10.6 & -7:10:46) &   ?     & --     & no obs          &  --   &  --  &  --   & ?         &  --        & --   &  --   &  1.0    &  --  & 150      &    -- \\ %Bow shock \\ 

131  & 5:39:57.4 & -7:29:33  & IRS~131 & 2.40   & no obs          &  --   &  --  &  --   & MMS~131   & 36x22(152) & 0.88 &  0.13 &  6.5    &  --  & 73       &    -- \\ %Emission NE of source\\
132  &(5:39:10.3 & -7:39:39) &   ?     & --     & no obs          &  --   &  --  &  --   & ?         &  --        & --   &  --   &$\sim$20 &  15  & $\sim$60 &    -- \\ %SMZ~76 counterflow? \\ 
133  & 5:39:34.3 & -7:26:11  & IRS~133?& 0.06   & no obs          &  --   &  --  &  --   &undetected &  --        & --   &  --   &$\sim$9.5&   9  & 170      &    -- \\ %Twisting chain of knots\\
134  &(5:41:06.2 & -8:00:14) &   ?     & --     & no obs          &  --   &  --  &  --   & no obs    &  --        & --   &  --   & $>$0.5  &  --  & $\sim$35 &    -- \\ %Compact arcs/bows \\
135  &(5:40:46.4 & -8:04:36) &   ?     & --     & no obs          &  --   &  --  &  --   & no obs    &  --        & --   &  --   &  2.5    &  --  & 90       &    -- \\ %Collimated, knotty jet\\ 
136  & 5:41:02.0 & -8:06:02  & IRS~136 & 1.18   & no obs          &  --   &  --  &  --   & no obs    &  --        & --   &  --   &  1.5    &  12  & 115      &    -- \\ %Faint filaments \\
137  & 5:41:01.7 & -8:06:45  & IRS~137 & 1.64   & no obs          &  --   &  --  &  --   & no obs    &  --        & --   &  --   &  1.5    &  15  & 60       &    -- \\ %Faint emission \\
138  & 5:40:48.8 & -8:06:57  & IRS~138 & 1.99   & no obs          &  --   &  --  &  --   & no obs    &  --        & --   &  --   & 13.4    &  45  & 130      &    -- \\ %Large bipolar jet/bows \\
139  &(5:41:23.9 & -8:12:47) &   ?     & --     & no obs          &  --   &  --  &  --   & no obs    &  --        & --   &  --   &$\sim$0.5&  --  & $\sim$80 &    -- \\ %Faint filament \\
140  &(5:42:31.9 & -8:01:14) &   ?     & --     & no obs          &  --   &  --  &  --   & no obs    &  --        & --   &  --   &  --     &  --  & --       &    -- \\ %Bright fingers \\ 

141  & 5:41:25.3 & -8:05:55  & IRS~141 & 0.29   & no obs          &  --   &  --  &  --   & no obs    &  --        & --   &  --   &  0.5    &  28  & 90       &    -- \\ %Faint bow nr source \\
142  &(5:39:42.4 & -8:02:58) &   ?     & --     & no obs          &  --   &  --  &  --   & no obs    &  --        & --   &  --   &  --     &  --  & --       &    -- \\ %Small group of knots \\
143  & 5:41:35.4 & -8:08:22  & IRS~143 & 1.64   & no obs          &  --   &  --  &  --   & no obs    &  --        & --   &  --   & $\sim$5 &  --  & $\sim$160&    -- \\ %Very faint emission \\
 
           %\noalign{\smallskip}
            \hline
           \noalign{\smallskip}
        \end{tabular}

\end{sidewaystable}

%%%%%%%%%%%%%%%%%%%%%%%%%%%%%%%%%%%%%%%%%%%%%%%%%%%%%%%%%%%%%%%%%%%%%%%%%%%%
%%%%%%%%%%%%%%%%%%%%%%%%%%%%%%%%%%%%%%%%%%%%%%%%%%%%%%%%%%%%%%%%%%%%%%%%%%%%
%%%%%%%%%%%%%%%%%%%%%%%%%%%%%%%%%%%%%%%%%%%%%%%%%%%%%%%%%%%%%%%%%%%%%%%%%%%%
%%%%%%%%%%%%%%%%%%%%%%%%%%%%%%%%%%%%%%%%%%%%%%%%%%%%%%%%%%%%%%%%%%%%%%%%%%%%
%%%%%%%%%%%%%%%%%%%%%%%%%%%%%%%%%%%%%%%%%%%%%%%%%%%%%%%%%%%%%%%%%%%%%%%%%%%%
%%%%%%%%%%%%%%%%%%%%%%%%%%%%%%%%%%%%%%%%%%%%%%%%%%%%%%%%%%%%%%%%%%%%%%%%%%%%
%%%%%%%%%%%%%%%%%%%%%%%%%%%%%%%%%%%%%%%%%%%%%%%%%%%%%%%%%%%%%%%%%%%%%%%%%%%%
%%%%%%%%%%%%%%%%%%%%%%%%%%%%%%%%%%%%%%%%%%%%%%%%%%%%%%%%%%%%%%%%%%%%%%%%%%%%
%%%%%%%%%%%%%%%%%%%%%%%%%%%%%%%%%%%%%%%%%%%%%%%%%%%%%%%%%%%%%%%%%%%%%%%%%%%%

\clearpage

\appendix

\section[]{Description of the H$_2$ outflows in Orion~A}

{\bf For a description of the new DFS outflows, and images of each flow,
please visit http://www.jach.hawaii.edu/$\sim$cdavis/}

\section{Proper motion measurements}

{\bf For a description of the proper motion measurements and
images of each flow with PM vectors superimposed, please visit
http://www.jach.hawaii.edu/$\sim$cdavis/}

\end{document}